\documentclass[12pt]{article}
\pdfoutput=1
\usepackage{amsfonts}
\usepackage{amssymb}
\usepackage{amsmath}
\usepackage{amstext}
\usepackage{graphicx}
\usepackage{epsfig}
\usepackage{verbatim}
\usepackage{fancyhdr}
\usepackage{fancybox}
\usepackage{color}
\usepackage{ulem,bbold}
\usepackage{enumitem}
\usepackage{subfigure}
\usepackage{bbm}
\usepackage{parskip}
\usepackage{latexsym}
\usepackage{theorem}
\usepackage{dsfont}
\usepackage{wasysym}
\usepackage[numbers,sort&compress]{natbib}

\newcommand{\Comment}[1]{{}}
\definecolor{MyDarkBlue}{rgb}{0.15,0.15,0.45}
\usepackage[linktocpage=true]{hyperref}
\hypersetup{
colorlinks=true,
citecolor=MyDarkBlue,
linkcolor=MyDarkBlue,
urlcolor=MyDarkBlue,
}

\newcommand{\bea}{\begin{eqnarray}}  
\newcommand{\eea}{\end{eqnarray}}

\def\nn{\nonumber}

\def\nl{\nonumber \\ &}
\def\nc{\, , \qquad}
\def\gt{\rightarrow}

\def\qq{\qquad}
\def\be#1\ee{\begin{align}#1\end{align}}
\def\del{\partial}
\def\grad{\nabla}
\def\({\left(}
\def\){\right)}
\def\[{\left[}
\def\]{\right]}

\def\half{\frac{1}{2}}

\def\td{\tilde{\d}}

\def\dt{\mathrm{d} t \,}

\def\dcx{\mathrm{d}^3 x \,}
\def\dcy{\mathrm{d}^3 y \,}

\def\ddx{\mathrm{d}^d x \,}
\def\ddy{\mathrm{d}^d y \,}
\def\ddz{\mathrm{d}^d z \,}
\def\H{\mathcal{H}}
\def\I{\mathcal{I}}
\def\Ib{\overline{\mathcal{I}}}

\def\J{\mathcal{J}}
\def\K{\mathcal{K}}
\def\M{\mathcal{M}}
\def\N{\mathcal{N}}

\def\d{\delta}
\def\e{\varepsilon}

\def\l{\lambda}

\def\w{\omega}

\def\W{\Omega}
\def\L{\Lambda}

\def\gradt{\tilde{\grad}}
\def\Ht{\tilde{\H}}

\def\Rt{\tilde{R}}

\def\pW{\pi_{\Omega}}
\def\pt{\tilde{\pi}}

\def\th{\tilde{h}}

\begin{document}

\begin{center}
{\Large \bf{On the Origin of Gravitational Lorentz Covariance}}
\end{center}

\vspace{1truecm}

\thispagestyle{empty} \centerline{
{\large  Justin Khoury${}^{a}$, Godfrey E. J. Miller${}^{b}$, and Andrew J. Tolley${}^{c}$}}

\vspace{1cm}

\centerline{{\it ${}^a$
Center for Particle Cosmology, Department of Physics \& Astronomy, University of Pennsylvania}}
\centerline{{\it 209 South 33rd Street, Philadelphia, PA 19104}}

\vspace{1cm}

\centerline{{\it ${}^b$
Princeton Consultants, Inc.}}
\centerline{{\it 2 Research Way, Princeton, NJ 08540}}

\vspace{1cm}

\centerline{{\it ${}^c$
Department of Physics, Case Western Reserve University}}
\centerline{{\it 10900 Euclid Ave, Cleveland, OH 44106}}

\vspace{1cm}

\begin{abstract}
We provide evidence that general relativity is the unique spatially covariant effective field theory of the transverse, traceless graviton degrees of freedom.  The Lorentz covariance of general relativity, having not been assumed in our analysis, is thus plausibly interpreted as an {\it accidental} or {\it emergent} symmetry of the gravitational sector.
\end{abstract}


\newpage
\setcounter{page}{1}

\section{Introduction}
\label{introduction}

Lorentz covariance is a central pillar of the modern field-theoretic interpretation of general relativity (GR).  From this point of view, GR is no more and no less than the unique Lorentz covariant theory of an interacting massless spin-2 particle~\cite{Feynman:1996kb,Weinberg:1965rz,Deser:1969wk}. In this paper, we show that GR can be derived without assuming Lorentz covariance.  Our approach relies on the weaker assumption of spatial covariance, within the context of the effective field theory of the transverse, traceless graviton degrees of freedom developed in~\cite{Khoury:2011ay}.

In canonical form, GR is a theory of a spatial metric $h_{i j}$ subject to first class\footnote{First class constraints are constraints that all commute with each other.} constraints $\H_\mu$.  Each constraint both generates a space-time gauge symmetry and eliminates a physical degree of freedom.  The Hamiltonian constraint $\H_0$ generates local time reparameterizations and eliminates the scalar polarization of the graviton; the momentum constraints $\H_i$ generate spatial diffeomorphisms and eliminate the longitudinal polarizations of the graviton.  In $3+1$ dimensions, the spatial metric has six components $h_{i j}$ subject to four constraints $\H_\mu$, so the graviton of general relativity has $6-4=2$ transverse, traceless polarizations.\footnote{See~\cite{Khoury:2011ay} for a comprehensive analysis of general relativity as a constrained field theory.}

In this framework, Lorentz covariance arises because the constraints $\H_\mu$ obey the Dirac algebra~\cite{Dirac:Lectures,Barbour:2000qg}, the algebra of the deformations of a spacelike hypersurface embedded in a Lorentzian space-time manifold~\cite{Teitelboim:1972vw}.  The project of relaxing the assumption of Lorentz covariance without introducing new gravitational degrees of freedom has been the subject of much research, and there are many directions one can take.  One can, for example, replace the Hamiltonian constraint with a restriction on the form of the Lagrangian~\cite{Barbour:2000qg}.  In the ultralocal truncation of GR, the Hamiltonian constraint still removes the scalar graviton polarization, but the constraints no longer satisfy the Dirac algebra~\cite{Farkas:2010dw}.  In the theory of Shape Dynamics, the Hamiltonian constraint is replaced by a constraint that generates volume preserving conformal transformations~\cite{Barbour:2011dn,Gomes:2011zj}.  In the generally covariant version of Ho\v{r}ava-Lifshitz gravity, additional fields and constraints are added to the theory to obtain the desired number of degrees of freedom~\cite{Horava:2010zj}.

Our method for relaxing the Hamiltonian constraint without introducing new degrees of freedom is to reduce the number of fields in the theory.  Our approach is inspired by a technique of conformal geometry, specifically the equivalence class construction underlying the notion of a conformal metric.  Any two metrics $h_{i j}$ and $h_{i j}'$ belong to the same conformal equivalence class if there exists a scalar function $\W$ such that $h_{i j}' = e^{\W} \cdot h_{i j}$. Each conformal class defines a geometry up to local changes of scale, and thereby defines a conformal metric.

Each conformal class also serves to define a unique {\it unit-determinant} metric.  To see this, consider an arbitrary representative metric $h_{i j}$ drawn from a conformal equivalence class.  Following the convention of~\cite{Farkas:2010dw}, in $d$ spatial dimensions we define
\be
\th_{i j} \equiv e^{-\W} h_{i j}
\nc
\W \equiv \frac{1}{d} \log h
\, .
\label{thphi}
\ee
By definition, $h_{i j}$ and $\th_{i j}$ belong to the same conformal equivalence class, and the metric $\th_{i j}$ has unit determinant.  Moreover, the metric $\th_{i j}$ so defined depends only on the choice of conformal class, not on the choice of representative within the class.  We have therefore established a one-to-one correspondence between conformal metrics and unit-determinant metrics.

Instead of thinking about gravity in terms of a spatial metric $h_{i j}$ subject to a system of constraints that kills its conformal mode, we will think of gravity in terms of a unit-determinant spatial metric $\th_{i j}$ that already lacks an independent conformal mode.  This allows us to remove the scalar polarization of the graviton while remaining agnostic about precisely how and why this polarization is absent.  Since it has one fewer component to begin with, a spatially covariant unit-determinant metric $\th_{i j}$ has the same number of degrees of freedom as a space-time covariant metric $h_{i j}$.  For example, in $3+1$ dimensions a unit-determinant spatial metric $\th_{i j}$ has five components, so subjecting such a metric to three momentum constraints $\Ht_i$ yields a graviton with $5-3=2$ transverse, traceless polarizations.  Insofar as we consider metrics with unit conformal factor, our proposal is similar in spirit to unimodular gravity~\cite{Henneaux:1989zc}.

By construction, spatially covariant theories of a unit-determinant metric describe the same transverse, traceless graviton polarizations as GR.  Though the kinematical state space is essentially the same, in principle the dynamical evolution of the graviton degrees of freedom could differ dramatically.  However, as we will see, demanding a consistent algebra and evolution for the momentum constraints singles out GR as the unique spatially covariant effective field theory of a transverse, traceless graviton.

The proof is quite technical in detail, but its outline is simple and can easily be summarized in a few steps. Our starting point, described in Sec.~\ref{action}, is the general canonical action in $d$ spatial dimensions
\be
S = \int \dt \ddx \( \dot{\th}_{i j} \pt^{i j} - \pi_H - N^i \Ht_i \)
\, ,
\label{SCGactionintro}
\ee
describing a unit-determinant metric $\th_{i j}$ and its traceless conjugate momentum $\pt^{i j}$. The scalar function $\pi_H$ is the physical Hamiltonian density. Spatial covariance is enforced
by the $\Ht_i$ momentum constraints, whose general form includes a second scalar function $\pi_K$:
\be
\Ht_i \equiv - 2 \th_{i j} \gradt_k \pt^{j k} - \gradt_i \pi_K
\, .
\label{kconformintro}
\ee
Here $\tilde{\nabla}$ denotes the covariant derivative with respect to $\th_{i j}$. Our set-up therefore depends on two functions, the physical Hamiltonian density $\pi_H$ and the momentum constraint density $\pi_K$,
which {\it a priori} are arbitrary functions of $t$, the phase space variables  $(\th_{i j}, \pt^{i j})$, and the spatial gradient operator $\partial_i$. In this language, GR corresponds to a particular form for $\pi_H$ and $\pi_K$, namely:
\be
\pi_H^{\rm GR} = - \dot{\W}(t) \pW
\nc
\pi_K^{\rm GR} = \frac{2}{d}  \pW
\, ,
\label{GRHKintro}
\ee
where
\be
\pW \equiv \pm \sqrt{d (d-1)} \sqrt{\phi(2) - \Rt \, e^{(d-1)\W(t) } + 2 \L e^{d\W(t)} }
\, .
\label{GRPWintro}
\ee
Here $\phi(2) \equiv \tilde{h}_{ik}\tilde{h}_{j \ell} \tilde{\pi}^{ij}\tilde{\pi}^{k\ell}$, $\Lambda$ is the cosmological constant, and $\Omega(t)$ is a monotonic function of time. For our general class of gravity theories, our only simplifying assumption about the form of $\pi_H$ and $\pi_K$ is that they depend on spatial gradients only through the Ricci scalar $\tilde{R}$. 

The allowed form for $\pi_H$ and $\pi_K$ is constrained by two considerations. First, in order to generate a consistent gauge symmetry, the momentum constraints $\Ht_i$ must be first-class under the action of the Poisson bracket,\footnote{The Poisson bracket on the phase space $( \th_{i j}, \pt^{i j} )$ is defined to be
\be
\{ A, B \} \equiv \int \ddx \( \frac{\d A}{\d \th_{r s}(x)} \frac{\d B}{\d \pt^{r s}(x)} - \frac{\d A}{\d \pt^{r s}(x)} \frac{\d B}{\d \th_{r s}(x)}\)
.
\ee
}
\be
\{ \Ht_i(x) , \Ht_a(y) \} \sim 0
\, .
\label{algebraconintro}
\ee
where $\sim$ denotes {\it weak equality}, {\it i.e.}, equality after the imposition of all constraints. The condition~\eqref{algebraconintro} means that the algebra of the constraints is {\it closed}, in the sense that the Poisson bracket of any two constraints is proportional to the constraints. The closure condition~(\ref{algebraconintro}) is studied in Sec.~\ref{closure}. After a considerable amount of algebra, we show how this translates into a (strong) condition on the form of $\pi_K$: either $\pi_K$ is an arbitrary ultralocal function,\footnote{By ultralocal, we mean that $\pi_K$ can depend on $t$ and the phase space variables  $\th_{i j}, \pt^{i j}$, but {\it not} on their spatial gradients.} or $\pi_K$ must be of the form
\be
\pi_K = \pm 2 \sqrt{\frac{(d-1)}{d}}\sqrt{\phi(2) + c_1(t) \Rt + c_2(t)}
\, ,
\label{pkrintro}
\ee
where $c_1(t)$ and $c_2(t)$ are arbitrary functions of time. Comparing with~(\ref{GRHKintro}) and~(\ref{GRPWintro}) above, we see that this is of the same form as the GR result, except for the more general time-dependence. 

The second constraint on the allowed form for $\pi_H$ and $\pi_K$ comes from demanding that the momentum constraints $\Ht_i$ be preserved under time evolution:
\be
\dot{\Ht}_i \sim 0
\, .
\ee
This condition is studied in Sec.~\ref{evolution}. After some algebra, this translates into a (strong) condition on the form of the physical Hamiltonian density $\pi_H$: in the ultra-local case, $\pi_H$ and $\pi_K$ are related by a differential equation --- see~(\ref{RGpiH}); in the case where $\pi_K$ takes the square-root form~(\ref{pkrintro}), $\pi_H$ is also constrained to take a square-root form, and the functions $c_1$, $c_2$ are related to one another by $c_2 = 2\lambda |c_1|^{d/(d-1)}$. Specifically, in this latter case the result is
\be
\pi_H = - \dot{\w}(t)\pi_\w
\nc
\pi_K = \frac{2}{d} \pi_\w
\, ,
\ee
where
\be
\pi_\w  \equiv \pm \sqrt{d (d-1)} \sqrt{\phi(2) \pm \Rt \, e^{(d-1)\w(t) } + 2 \l e^{d\w(t)}}\,,
\ee
with $\w(t) \equiv \frac{1}{d-1} \log |c_1(t)|$. With the replacements $\omega \rightarrow \Omega, \lambda \rightarrow \Lambda$, this becomes identical to the GR result given by~\eqref{GRHKintro} and~\eqref{GRPWintro}. 

To summarize, the combined requirements of closure of the constraints and consistency under time evolution together single out GR as the unique effective field theory of the graviton degrees of freedom. We conclude in Sec.~\ref{conclusion} with a discussion of the possible implications of these results for our understanding of space-time symmetry in gravitational theory.

\section{Spatially-Covariant, Transverse, Traceless Gravitons}
\label{action}

In this section, we construct the spatially covariant effective action of the transverse, traceless graviton polarizations.  We describe these degrees of freedom using a unit-determinant metric $\th_{i j}$ subject to momentum constraints $\Ht_i$. Before proceeding with the nonlinear analysis in Sec.~\ref{dynamicsGR}, we first discuss the essence of our approach from a perturbative point of view.

\subsection{Perturbative graviton}

As we have already discussed, GR is the unique Lorentz invariant low energy theory describing a single interacting massless spin-two particle, i.e. an interacting spin-two representation of the Poincar\'e group~\cite{Feynman:1996kb,Weinberg:1965rz,Deser:1969wk}. Since it is our intention to see if it is possible to construct an interacting theory of a transverse, traceless graviton without assuming Lorentz invariance from the outset, we can already ask at the level of the free graviton theory what extra freedom we have. Assuming only spatial rotations and space translations, and denoting the two transverse traceless tensor degrees of freedom collectively as $h_{TT}$, which we implicitly assume to be living on a spatially flat FRW background, the action for the free graviton theory must take the form at lowest order in spatial derivatives
\be
S_{\rm graviton} = \int \dt \int \ddx  \frac{1}{4}  \left( A^2(t)(\dot h_{TT})^2 - A^2(t) c_s^2(t) (\partial_i h_{TT})^2 - A^2(t)M(t)^2 h_{TT}^2 \right) .
\ee
Here $M(t)$ denotes an effective mass which may arise in a cosmological context, $A(t)$ denotes the overall normalization of the graviton fluctuations, and $c_s(t)$ determines the speed of propagation which also may in principle be time-dependent. This action may also be supplemented by higher spatial gradient terms which are not forbidden by symmetry, e.g. $(\nabla^2 h_{TT})^2$, but by assumption not higher time derivatives as these would lead to additional propagating degrees of freedom which would in general be ghosts. Focusing on the lowest order derivative terms, in the absence of a specified coupling to matter, we are free to perform a time reparameterization $t \rightarrow f(t)$ and use this to set $c_s(t)=1$, after an appropriate change of $A(t)$ and $M(t)$. Furthermore we can perform a field redefinition $h_{TT} \rightarrow h_{TT} a(t)/A(t)$ to make the graviton normalization conventional and consistent with the background FRW geometry defined by the scale factor $a(t)$, at the price of a redefinition of the effective mass $M(t) \rightarrow \tilde M(t)$. After these transformations the free graviton action will take the form
\be
\label{graviton}
S_{\rm graviton} = \int \dt \int \ddx  \frac{1}{4}  \left( a^2(t)(\dot h_{TT})^2 - a^2(t) (\partial_i h_{TT})^2 - a^2(t)\tilde M(t)^2 h_{TT}^2 \right) .
\ee
There is however no additional freedom to remove the effective mass term $\tilde M(t)^2$. Furthermore it makes no sense to restrict to a massless graviton since `mass' is only uniquely defined in Minkowski space-time in terms of the Casimir $P_{\mu}P^{\mu}$ of the Poincar\'e group and as we have seen in an FRW geometry a time-dependent mass term can be generated from the normalization of the wavefunction. 
Consequently we see that in the absence of Poincar\'e/Lorentz invariance (or de Sitter), even in the free graviton theory we have a single free function of time $\tilde M(t)$ which encodes potential departures from GR. As we add higher spatial gradient terms we have more free functions of time even in the free theory, e.g. $C(t)(\nabla^2 h_{TT})^2$. In short, the question we would like to address in this paper is whether there exists a nonlinear completion of the free theory Eq.~(\ref{graviton}) which preserves only the two propagating transverse traceless degrees of freedom nonlinearly. We now proceed to the nonlinear theory, and begin with the identification of the transverse traceless degree of freedom in GR.

\subsection{The dynamics of GR}

\label{dynamicsGR}

In terms of the ADM~\cite{Arnowitt:1962hi} variables,
\be
{\rm d}s^2 = g_{\mu \nu} {\rm d}x^{\mu} {\rm d}x^{\nu} \equiv - N^2 {\rm d}t^2 + h_{i j} \( {\rm d}x^i + N^i {\rm d}t \) \( {\rm d}x^j + N^j {\rm d}t \)\,,
\ee
the action of GR in canonical form is
\be
S = \int \dt \ddx \( \dot{h}_{i j} \pi^{i j} - N^\mu \H_\mu \)
\, ,
\label{cGR}
\ee
where $d$ is the number of spatial dimensions, $\pi^{i j}$ is the canonical momentum conjugate to $h_{i j}$, and $N^0 \equiv N$.  The lapse scalar $N$ appears in~\eqref{cGR} as a Lagrange multiplier which enforces the vanishing of the Hamiltonian constraint
\be
\H_0 \equiv \frac{1}{\sqrt{h}} \( h_{a c} h_{b d} - \frac{1}{(d-1)} h_{a b} h_{c d} \) \pi^{a b} \pi^{c d} - \sqrt{h} \( R - 2 \L \)
\,,
\label{H0}
\ee
where $h$ is the determinant of $h_{i j}$, $R$ is the Ricci scalar of $h_{i j}$, and $\L$ is the cosmological constant.  Similarly, the components of the shift vector $N^i$ appear in~\eqref{cGR} as Lagrange multipliers which enforce the vanishing of the momentum constraints
\be
\H_i \equiv - 2 h_{i j} \grad_k \pi^{j k}
\, ,
\label{Hi}
\ee
where the symbol $\grad_i$ denotes the covariant derivative with respect to $h_{i j}$.  General relativity is thus properly understood as a constrained field theory.\footnote{See~\cite{Henneaux:1992ig,Hanson:1976cn} for pedagogical treatments of the general theory of constrained systems.} Up to a sign, the Hamiltonian constraint~\eqref{H0} fixes the trace of the momentum tensor ($h_{i j} \pi^{i j}$), thereby eliminating the scalar polarization of the graviton.  The momentum constraints fix the divergence of the momentum tensor ($\grad_j \pi^{i j}$), eliminating the longitudinal polarizations of the graviton.  Accordingly, the graviton of GR has only transverse, traceless polarizations.

As discussed in~\cite{Khoury:2011ay}, the trace of the momentum tensor is conjugate to the conformal part of the spatial metric, so the Hamiltonian constraint renders the conformal factor essentially non-dynamical.  To see this, decompose the spatial metric into a conformal part $e^{\W}$ and a unit-determinant part $\th_{i j}$,
\be
h_{i j} = e^{\W} \th_{i j}
\, .
\label{condec}
\ee
Inserting this decomposition into the first term of the action~\eqref{cGR} yields
\be
\dot{h}_{i j} \pi^{i j} = \dot{\W} \( h_{i j} \pi^{i j} \) + \dot{\th}_{i j} \( e^{\W} \pi^{i j} \)
\, .
\label{firstterm}
\ee
Inspecting the coefficients of $\dot{\W}$ and $\dot{\th}_{i j}$ in this equation allows us to extend the conformal decomposition to the momentum tensor.  First, it is obvious that the conformal mode $\W$ is conjugate to the trace of the momentum tensor.  Second, since $\dot{\th}_{i j}$ is traceless, the momentum conjugate to the unit-determinant metric $\th_{i j}$ is proportional to the traceless part of the momentum tensor.\footnote{The tracelessness of $\dot{\th}_{i j}$ follows from the identity $\d \th = \th \cdot \th^{i j} \d \th_{i j}$ and the condition $\th \equiv 1$.}  We therefore define
\be
\pW \equiv h_{i j} \pi^{i j} \,;
\qquad
\pt^{i j} \equiv e^{\W} \( \pi^{i j} - \frac{1}{d} h^{i j} h_{a b} \pi^{a b} \)
\,,
\ee
where $\pt^{i j}$ is manifestly traceless.  The momentum tensor $\pi^{i j}$ admits the trace decomposition
\be
\pi^{i j} = e^{-\W} \pt^{i j} + \frac{1}{d} h^{i j} \pW
\, .
\label{tracedec}
\ee
Meanwhile,~\eqref{firstterm} simplifies to
\be
\dot{h}_{i j} \pi^{i j} = \dot{\W} \pW + \dot{\th}_{i j} \pt^{i j}
\, ,
\ee
so the spatial metric phase space $(h_{i j},\pi^{i j})$ splits naturally into a conformal part $(\W, \pW)$ and a unit-determinant part $(\th_{i j},\pt^{i j})$.  Schematically,
\be
(h_{i j},\pi^{i j}) \gt (\W, \pW) \, , \, (\th_{i j},\pt^{i j})
\, .
\ee
Having performed this phase space decomposition, we will now examine how the Hamiltonian constraint $\H_0$ fixes the magnitude of $\pW$ and the momentum constraints $\H_i$ fix the divergence of $\pt^{i j}$.  Substituting the trace decomposition~\eqref{tracedec} into the Hamiltonian constraint~\eqref{H0} yields
\be
\H_0 \equiv \frac{1}{\sqrt{h}} \( \th_{a c} \th_{b d} \pt^{a b} \pt^{c d} - \frac{\pW^2}{d(d-1)} \)  - \sqrt{h} \( R - 2 \L \)
.
\ee
Similarly, the momentum constraints~\eqref{Hi} can be written
\be
\H_i = - 2 \th_{i j} \grad_k \pt^{j k} - \frac{2}{d} \grad_i \pW
\, .
\label{traceHi}
\ee
Up to a sign, the Hamiltonian constraint $\H_0$ forces $\pW$ to be a function of the remaining phase space variables.  Since the momentum conjugate to $\W$ is almost completely fixed by the Hamiltonian constraint, $\W$ does not correspond to a true, propagating physical degree of freedom.  At the same time, the $\H_i$'s constrain the longitudinal part of $\pt^{i j}$ to be proportional to the gradient of $\pW$.  The only momentum variables which can be freely specified as initial data are the transverse components of the traceless momentum tensor $\pt^{i j}$.

\subsection{Our set-up}
To relax Lorentz covariance to spatial covariance, we drop the Hamiltonian constraint $\H_0$.  To avoid a scalar graviton polarization, we also drop the conformal part of the phase space, $(\W , \pW)$.  We therefore choose our phase space to consist of a unit-determinant metric $\th_{i j}$ and a traceless conjugate momentum $\pt^{i j}$ subject to momentum constraints $\Ht_i$ that enforce spatial covariance.  The most general canonical action of this kind takes the form
\be
S = \int \dt \ddx \( \dot{\th}_{i j} \pt^{i j} - \pi_H - N^i \Ht_i \)
\, ,
\label{SCGaction}
\ee
where $\pi_H$ is the physical Hamiltonian density.  Note that $\th_{i j}$ has $(d+2)(d-1)/2$ components, so the spatial dimension must obey $d \geq 2$.

In the action~\eqref{SCGaction}, the Lagrange multipliers $N^i$ enforce the vanishing of the momentum constraints $\Ht_i$.  To determine the form of the $\Ht_i$'s, we will adapt the $\H_i$'s from~\eqref{traceHi} to the reduced phase space $(\th_{i j},\pt^{i j})$.  This ensures that our action is general enough to represent all spatially covariant theories of a spatial metric $h_{i j}$ for which the conformal mode happens to be non-dynamical.  In terms of the covariant derivative $\gradt_i$ with respect to the metric $\th_{i j}$, equation~\eqref{traceHi} becomes
\be
\H_i = - 2 \th_{i j} \gradt_k \pt^{j k}
- \( d + 2 \) \cdot \th_{i j} \pt^{j k} \gradt_k \W
- \frac{2}{d} \gradt_i \pW
\, .
\label{expandHi}
\ee
To simplify the form of the corresponding $\Ht_i$'s, we make use of the fact that we are treating gravity as an effective field theory to eliminate $\W$ as an independent function.

To treat gravity as the effective field theory of $\th_{i j}$ and $\pt^{i j}$, we must choose a background about which to define the action.  We wish our background solution to respect the global spatial symmetries of the theory, so we look for maximally symmetric background field configurations $\th_{i j}^0$ and $\pt^{i j}_0$.  As is well known from FRW cosmology, maximally symmetric spatial metrics can be flat, open, or closed.  However, the fixed conformal factor of a unit determinant metric rules out the open and closed possibilities, so the unique maximally symmetric unit-determinant spatial metric is
\be
\th_{i j}^0 = \d_{i j}
\, .
\ee
Maximally symmetric rank-2 tensors living in a maximally symmetric space must either vanish identically or be proportional to the metric itself.  The momentum tensor is traceless, so it cannot be proportional to the metric.  It must therefore vanish on the maximally symmetric background,
\be
\pt^{i j}_0 = 0
\, .
\ee
Next consider the behavior of the two scalar functions $\W$ and $\pW$ on this background.  To be consistent with spatial homogeneity and isotropy, the background scalars $\W_0$ and $\pW^0$ can depend only on time,
\be
\W_0(t)
\nc
\pW^0(t)
\, .
\ee
Now consider deformations from the maximally symmetric background.  Up to this point, we do not have an independent definition of time, so we can use our coordinate freedom to choose one of the scalar functions, $\W$ or $\pW$, to serve as a time coordinate.  Choosing $\pW$ as the definition of time corresponds to ``constant mean curvature'' gauge, while choosing $\W$ as the definition of time corresponds to using redshift to define ``look-back time.''  Either approach is consistent so long as the chosen function evolves monotonically.  On generic backgrounds, such a definition of time can be applied locally, but on cosmological backgrounds such a definition can be applied globally.

In this paper, we choose $\W$ to serve as our time coordinate; inverting this relation allows us to write $\W$ as a function of time.  Since $\W$ is a function of time, spatial gradients of $\W$ vanish.  Applying this observation to equation~\eqref{expandHi}, we define
\be
\Ht_i \equiv - 2 \th_{i j} \gradt_k \pt^{j k} - \gradt_i \pi_K
\, ,
\label{kconform}
\ee
where $\pi_K$ is an arbitrary function of time $t$, the phase space variable $\th_{i j}$ and $\pt^{i j}$, and the spatial gradient operator $\del_i$.  This completes our specification of the action~\eqref{SCGaction}, which we reproduce here for convenience:
\be
S = \int \dt \ddx \( \dot{\th}_{i j} \pt^{i j} - \pi_H - N^i \Ht_i \) \, .
\label{HKaction}
\ee
The action~\eqref{HKaction} depends on two unspecified scalar functions: the physical Hamiltonian density $\pi_H$ and the momentum constraint density $\pi_K$\footnote{We shall throughout make the additional assumption that as in GR, these functions are local in the sense that they admit a well-defined derivative expansion in spatial derivatives. In \cite{Bellorin:2013zbp}, it is argued that a Lorentz violating theory of only two degrees of freedom exists in the context of Horava-Lifshitz theory. However the framework of \cite{Bellorin:2013zbp} relies on the introduction of spatial derivatives for the lapse $N$ in the action. On integrating out $N$, to put the action on the present form Eq.~(\ref{HKaction}), nonlocal inverse Laplacians would be generated in the analogue of $\pi_H$. By contrast here we work with a strictly local form for the Hamiltonian.}.  In the enlarged phase space $(h_{i j}, \pt^{i j})$, the function $\pi_K$ would be interpreted as proportional to $\pW$.  As we mentioned before, the advantage of our approach is that we can remain agnostic about how the conformal mode is constrained.  By construction, gravitons in this theory lack a scalar polarization, and are therefore automatically traceless.  Assuming that the momentum constraint algebra closes properly (see Sec.~\ref{closure}) and that the equations of motion preserve the constraints (see Sec.~\ref{evolution}), the gravitons of this theory are guaranteed to have the correct number of polarizations.

\subsection{Building $\pi_H$ and $\pi_K$}

The action~\eqref{HKaction} depends on two scalar functions $\pi_H$ and $\pi_K$ which we will now build in full generality.  The most general scalars consistent with spatial covariance can depend explicitly on time $t$, and can be built out of three objects with spatial indices, namely the unit-determinant metric $\th_{i j}$, the traceless momentum $\pt^{i j}$, and the spatial gradient operator $\del_i$.  For convenience, the metric, momentum, and gradient can be organized into the four spatially-covariant tensorial objects
\be
\th_{i j}
\nc
\pt^{i j}
\nc
\Rt_{a b c d}
\nc
\gradt_i
\, ,
\ee
where $\Rt_{a b c d}$ is the Riemann tensor of $\th_{i j}$.  So far, $\pi_H$ and $\pi_K$ are unspecified scalar functions of time $t$ and the four tensorial building blocks.  To enumerate all the distinct scalars on which $\pi_H$ and $\pi_K$ will be allowed to depend, we again employ the logic of effective field theory.  In determining the relevance of a given scalar, each momentum tensor $\pt^{i j}$ is accompanied by a temporal expansion parameter $\e_t$, while each covariant derivative $\gradt_i$ is accompanied by a spatial expansion parameter $\e_s$.  For the effective field theory approach to be well-defined, we assume that $0 < \e_t < 1$ and $0 < \e_s < 1$.  The product of expansion parameters associated with a given scalar will be termed its {\it relevance coefficient}.

At zeroth order in $\e_s$, the possible scalars are {\it ultralocal} functions of the phase space variables, {\it i.e.}, functions constructed solely out of contractions of $\th_{i j}$ and $\pt^{i j}$ without any spatial gradients~\cite{Farkas:2010dw}.  The distinct fully connected ultralocal scalars can be thought of as chains of momentum tensors contracted into rings.  Define the scalar $\phi(n)$ to be the fully connected contraction of $n$ factors of $\pt^{i j}$, {\it i.e.},
\be
\phi(n) \equiv \th_{i j} \Pi^{i j}(n)
\nc
\Pi^{i j}(n+1) \equiv \pt^{i a} \th_{a b} \Pi^{b j}(n)
\nc
\Pi^{i j}(0) \equiv \th^{i j}
\, .
\ee
Since $\phi(0) = d$ and $\phi(1) = 0$, any ultralocal scalar will be some function of the $\phi(n)$'s with $n \geq 2$.  In terms of our expansion parameters,
\be
\phi(n) \sim \e_t^n \cdot \e_s^0
\, ,
\ee
so the relevance coefficient of $\phi(n)$ is simply $\e_t^n$.

As it turns out, there are no scalars of odd order in $\e_s$.  This is because the spatial gradient $\del_i$ is the only object in the theory with an odd number of indices, so any quantity of odd order in gradients will always have at least one free index.

Each scalar of second order in $\e_s$ is built by contracting factors of $\th_{i j}$ and $\pt^{i j}$ against one of the following three tensors,
\be
\Rt_{a b c d}
\nc
\gradt_a \gradt_b \pt^{i j}
\nc
\gradt_r \pt^{a b}  \gradt_s \pt^{i j} 
\, .
\ee
The leading two-derivative scalar is the Ricci scalar, given by $\Rt \equiv \th^{a c} \th^{b d} \Rt_{a b c d}$.  In terms of our expansion parameters,
\be
\Rt \sim \e_s^2
\, ,
\ee
so the relevance coefficient of $\Rt$ is $\e_s^2$.  All remaining scalars of second order in $\e_s$ are suppressed by additional factors of $\e_t$. The Ricci scalar thus represents the leading local correction to ultralocal physics.\footnote{Any other scalar at the same order in $\epsilon_s$ is necessarily higher-order in $\epsilon_t$. For instance, $\pt^{a c} \pt^{b d} \Rt_{a b c d} \sim \epsilon_t^2\epsilon_s^2$.}

To implement a consistent effective truncation of the general action, we introduce a truncation parameter $\d>0$ and exclude all scalars with a relevance coefficient less than or equal to $\d$.  In this paper, we wish to examine the leading local correction to ultralocal physics, so we assume $\e_s < \e_t$.  To ensure that the leading local scalar ({\it i.e.}, $\Rt$) is not excluded, we choose $\d < \e_s^2$.  To exclude local scalars of higher order in time derivatives, we choose $\d > \e_s^2 \e_t$; since $\e_t > \e_s$, this choice also excludes local scalars of higher order in spatial derivatives.

To determine which ultralocal scalars are not excluded, let $N$ momentarily denote the largest integer for which $\e_t^N > \d$.  The truncated action is then allowed to depend on the ultralocal scalars $\phi(2),\phi(3),\ldots,\phi(N)$.  In principle, however, $N$ could be arbitrarily large, so the action could depend on any given ultralocal scalar.  In practice, therefore, we will not restrict the possible dependence of the action on ultralocal scalars.

Following truncation, $\pi_H$ and $\pi_K$ could depend on any of the ultralocal scalars, but can depend on spatial gradients only through $\Rt$. In other words, 
\be
\pi_H = \pi_H( t , \phi(n), \Rt )
\nc
\pi_K = \pi_K( t , \phi(n) , \Rt )
\, ,
\label{hk}
\ee
where $\phi(n)$ stands collectively for $\phi(2),\phi(3), \ldots$

\subsection{Example: Spatially Covariant General Relativity}

As an example of our formalism, let us review how to cast the space-time covariant action~\eqref{cGR} of GR into the spatially covariant form~\eqref{HKaction}~\cite{Khoury:2011ay}. 
Following~\eqref{thphi}, we split the metric $h_{i j}$ into a conformal part $\W$ and a unit-determinant part $\th_{i j}$.  About cosmological backgrounds, we take the conformal factor $\W$ to be our measure of time.  Formally, we can use a Lagrange multiplier $\psi$ to enforce a gauge-fixing constraint $\chi \equiv \W- \W(t) = 0$, so that the new gauge-fixed action is
\be
S' = \int \dt \ddx \( \pi^{i j}\dot{h}_{i j} - N^{\mu} \H_{\mu} - \psi \chi \) .
\ee
The constraints $\H_0$ and $\chi$ are second class,\footnote{Second class constraints are constraints which do not commute under the action of the Poisson bracket.} so they can be solved explicitly to yield expressions for $\W$ and its conjugate momentum in terms of $t$, $\th_{i j}$, $\pt^{i j}$ and spatial derivatives.  Substituting these expressions back into the gauge-fixed action yields an action of the form~\eqref{HKaction}, with
\be
\pi_H = - \dot{\W}(t) \pW
\nc
\pi_K = \frac{2}{d}  \pW
\, ,
\label{GRHK}
\ee
and
\be
\pW = \pm \sqrt{d (d-1)} \sqrt{\phi(2) - \Rt \, e^{(d-1)\W(t) } + 2 \L e^{d\W(t)} }
\, .
\label{GRPW}
\ee
We verified in~\cite{Khoury:2011ay} that GR in spatially-covariant gauge is consistent as a constrained field theory.  In the next two sections, we will show that it is essentially the unique effective field theory emerging from the action~\eqref{HKaction}.

\section{Closure of the Constraint Algebra}
\label{closure}

To generate a consistent gauge symmetry, the momentum constraints $\Ht_i$ given by~\eqref{kconform} must be first-class under the action of the Poisson bracket,
\be
\{ \Ht_i(x) , \Ht_a(y) \} \sim 0
\, .
\label{algebracon}
\ee
The symbol $\sim$ denotes {\it weak equality}, which means equality after the imposition of all constraints.  The condition~\eqref{algebracon} means that the algebra of the constraints is {\it closed}, in the sense that the Poisson bracket between any two constraints yields an expression proportional to the constraints.

Satisfying the weak equality~\eqref{algebracon} will impose restrictions on the form of $\pi_K$.  As we will see, there are only two solutions to this condition: either $\pi_K$ must be an ultralocal function of $t$ and the phase space variables, or $\pi_K$ must take the GR form~\eqref{GRHK}.  However, the constraint redundancy implicit in the notion of weak equality makes it impossible to solve~\eqref{algebracon} directly for $\pi_K$. To proceed, we must first derive a {\it strong equation}, one that holds identically irrespective of any constraints.

\subsection{Deriving A Strong Equation}

We defer the computation of the Poisson bracket $\{ \Ht_i(x) , \Ht_a(y) \}$ to Appendix~\ref{algebra}.  The result is (see~\eqref{HtiHta} and~\eqref{Ik})
\be
\{ \Ht_i(x) , \Ht_a(y) \}
&
= \Ht_a(x) \del_{x^i} \d^3(x-y) - \Ht_i(y) \del_{y^a} \d^3(x-y)
\nl
+ \del_{x^i} \( - \I^k(x) \del_{x^k} \del_{x^a} \d^3(x-y) \)
- \del_{y^a} \( - \I^k(y) \del_{y^k} \del_{y^i} \d^3(x-y) \)
,
\label{momentumbracket}
\ee
where
\be
\I_k
&
=
\frac{\del \pi_K}{\del \Rt} \gradt^j \( \sum_{m = 2}^{\infty} m \frac{\del \pi_K}{\del \phi(m)} \Pi(m-1)_{b c} \td_{j k}^{b c} \)
\nl
- \( \sum_{m = 2}^{\infty} m \frac{\del \pi_K}{\del \phi(m)} \Pi(m-1)_{b c} \td_{j k}^{b c} \) \gradt^j \frac{\del \pi_K}{\del \Rt}
- \frac{2 (d-1)}{d} \gradt_k \frac{\del \pi_K}{\del \Rt}
\, .
\label{Ikup}
\ee
In the expression for $\I_k$, the symbol $\td^{i j}_{a b}$ denotes the traceless, symmetric Kronecker matrix,
\be
\td^{i j}_{a b} = \half \d^i_a \d^j_b + \half \d^j_a \d^i_b - \frac{1}{d} \th^{i j} \th_{a b}
\, .
\ee
Since $\Ht_i \sim 0$ by assumption,~\eqref{momentumbracket} agrees with condition~\eqref{algebracon} if and only if
\be
\I_k \sim 0
\, .
\label{weak}
\ee
The weak equality in~\eqref{weak} is the necessary and sufficient condition for the momentum constraints $\Ht_i$ to generate a consistent first class constraint algebra~\cite{Khoury:2011ay}.  To solve~\eqref{weak}, we will first eliminate the constraint redundancy, and thereby promote the weak equality to a strong equation.

Recall that the momentum constraints $\Ht_i$~\eqref{HKaction} fix the longitudinal part of the momentum tensor, $\gradt_j \pt^{i j}$, in terms of the gradient of a scalar function, $\gradt_i \pi_K$.  In particular, the momentum constraints enforce the weak equality
\be
\gradt_j \pt^{i j} \sim - \half \gradt^i \pi_K
\, .
\ee
We therefore search the definition of $\I_k$ in equation~\eqref{Ikup} for every instance of $\gradt_j \pt^{i j}$, {\it i.e.}, for every instance in which a derivative operator 1) acts on a momentum tensor and 2) is contracted against one of the indices of that tensor.  Next, we perform the substitution
\be
\gradt_j \pt^{i j} \gt - \half \gradt^i \pi_K
\, .
\label{sub}
\ee
After removing every instance of $\gradt_j \pt^{i j}$ in this manner, we can write $\I_k \sim \Ib_k$, where
\be
\Ib_k
&
\equiv
\frac{\del \pi_K}{\del \Rt} \sum_{m = 3}^{\infty} m \frac{\del \pi_K}{\del \phi(m)} \pt^{j a} \gradt_j \Pi(m-2)_{a k}
+ \frac{\del \pi_K}{\del \Rt} \( \sum_{m = 2}^{\infty} m \Pi(m-1)_{k j} \gradt^j \frac{\del \pi_K}{\del \phi(m)} \)
\nl
- \half \frac{\del \pi_K}{\del \Rt} \( \sum_{m = 2}^{\infty} m \frac{\del \pi_K}{\del \phi(m)} \Pi(m-2)_{k a} \) \gradt^a \pi_K
- \frac{1}{d} \frac{\del \pi_K}{\del \Rt} \gradt_k \( \sum_{m = 3}^{\infty} m \frac{\del \pi_K}{\del \phi(m)} \phi(m-1) \)
\nl
- \( \sum_{m = 2}^{\infty} m \frac{\del \pi_K}{\del \phi(m)} \Pi(m-1)_{j k} \) \gradt^j \frac{\del \pi_K}{\del \Rt}
+ \frac{1}{d} \( \sum_{m = 3}^{\infty} m \frac{\del \pi_K}{\del \phi(m)} \phi(m-1) \) \gradt_k \frac{\del \pi_K}{\del \Rt}
\nl
- \frac{2 (d-1)}{d} \gradt_k \frac{\del \pi_K}{\del \Rt}
\, .
\label{Ibk}
\ee
It follows immediately that $\Ib_k \sim 0$, but we can do better.  Since $\Ib_k$ does not contain any factors of $\gradt_j \pt^{i j}$ or $\Ht_i$, the weak equality $\Ib_k \sim 0$ implies the strong equation
\be
\Ib_k = 0
\, .
\label{strong}
\ee
This is the desired result. This strong equation is a necessary and sufficient condition for the constraints to generate a consistent first-class algebra, and it contains no constraint redundancy.  
In what follows, we will solve this equation to determine the allowed form of $\pi_K$.

\subsection{Solving The Strong Equation}

The simplest way to solve the strong equation~\eqref{strong} is for $\pi_K$ to be an {\it ultralocal} function. Ultralocal theories of gravity have interesting applications near cosmological singularities~\cite{Belinsky:1970ew,Henneaux:2007ej} and for studying cosmological perturbations on super-horizon scales~\cite{Salopek:1990mp}. By ultralocal, we mean that $\pi_K$ depends on $t$
and the phase space variables $\th_{i j}$ and $\pt^{i j}$, but not on their spatial gradients. In particular, $\del \pi_K/\del \Rt = 0$ in this case, and $\Ib_k$ vanishes identically. {\it Therefore, any ultralocal $\pi_K= \pi_K(t, \phi(n))$ satisfies~\eqref{strong}.} 

However, in the ultralocal limit, each point in space evolves like a separate universe, so considerations of causality demand that we allow for {\it local} coupling between adjacent points in space.\footnote{For more on ultralocal gravity and the separate universe approximation, see~\cite{Tolley:2008na}.}  To understand the propagation of local graviton degrees of freedom, it is of course necessary to go beyond the ultralocal limit. Consistent with our earlier assumptions, we allow $\pi_K$ to depend on spatial gradients only through the Ricci scalar.

In the local case, determining the form of $\pi_K$ is equivalent to solving the strong equation~\eqref{strong} assuming $\del \pi_K/\del \Rt \neq 0$.  The solution can be obtained in a straightforward manner analogous to the technique known as {\it separation of variables}: if one side of an equation depends on a variable $X$ and the other does not, then in fact neither side depends on $X$.  The left-hand-side of~\eqref{strong} is the vector quantity $\Ib_k$, which depends non-trivially on an infinite number of distinct phase space scalars and vectors; the right-hand-side of~\eqref{strong} is $0$, which does not depend on any phase space quantities.  Keep in mind that equation~\eqref{strong} cannot be solved by positing additional restrictions on the phase space scalars and vectors, because this would reduce the number of physical degrees of freedom.  Eliminating the dependence of $\Ib_k$ on a given phase space quantity therefore imposes a corresponding restriction on the form of $\pi_K$.  The most general $\pi_K$ consistent with all such restrictions is the general solution to~\eqref{strong}.  In practice, it is easiest to solve the strong equation iteratively by

1) picking a phase space quantity which appears in only a few terms of $\Ib_k$;

2) determining the corresponding restriction on $\pi_K$;

3) substituting the restriction back into the strong equation~\eqref{strong}, thereby simplifying it;

4) repeating until the strong equation is solved.

To quickly simplify~\eqref{strong}, consider the dependence of $\Ib_k$ on the phase space vectors
\be
\pt^{j a} \gradt_j \Pi(m-2)_{a k}
\nc
m \geq 3
\, .
\ee
As is clear from equation~\eqref{Ibk}, each such vector enters $\Ib_k$ with a scalar coefficient
\be
m \frac{\del \pi_K}{\del \Rt} \frac{\del \pi_K}{\del \phi(m)}
\, .
\ee
To eliminate the dependence of $\Ib_k$ on the vectors $\pt^{j a} \gradt_j \Pi(m-2)_{a k}$ without generating new constraints on phase space, each of these scalar coefficients must be made to vanish.  Since $\del \pi_K/\del \Rt \neq 0$, it must be the case that
\be
\frac{\del \pi_K}{\del \phi(m)} = 0
\nc
m \geq 3
\, .
\ee
In other words, the most general form for $\pi_K$ is
\be
\pi_K = \pi_K(t, \phi(2),\Rt)\,,
\ee
in which case the strong equation~\eqref{strong} reduces to
\be
0
&
=
2 \(
\frac{\del \pi_K}{\del \Rt} \frac{\del^2 \pi_K}{\del \phi(2)^2}
- \frac{\del \pi_K}{\del \phi(2)} \frac{\del^2 \pi_K}{\del \phi(2) \, \del \Rt}
\) \pt_k^{\ j} \gradt_j \phi(2) 
\nl
+ 2 \( \frac{\del \pi_K}{\del \Rt} \frac{\del^2 \pi_K}{\del \phi(2) \, \del \Rt}
- \frac{\del \pi_K}{\del \phi(2)} \frac{\del^2 \pi_K}{\del \Rt^2}
\) \pt_k^{\ j} \gradt_j \Rt
\nl
- \(
\frac{\del \pi_K}{\del \Rt} \( \frac{\del \pi_K}{\del \phi(2)} \)^2
+ \frac{2 (d-1)}{d} \frac{\del^2 \pi_K}{\del \phi(2) \, \del \Rt}
\) \gradt_k \phi(2)
\nl
- \( \frac{\del \pi_K}{\del \phi(2)} \( \frac{\del \pi_K}{\del \Rt} \)^2
+ \frac{2 (d-1)}{d} \frac{\del^2 \pi_K}{\del \Rt^2}
\) \gradt_k \Rt
\, .
\ee
For this equation to hold without introducing new constraints on the phase space variables, the coefficients of $\pt_k^{\ j} \gradt_j \phi(2)$, $\pt_k^{\ j} \gradt_j \Rt$, $\gradt_k \phi(2)$ and 
$\gradt_k \Rt$ must vanish independently. Specifically, the coefficients of the last two lines imply
\be
\frac{\del^2 \pi_K}{\del \phi(2) \, \del \Rt}
&=
\frac{-d}{2 (d-1)}
\( \frac{\del \pi_K}{\del \phi(2)} \)^2
\frac{\del \pi_K}{\del \Rt}
\label{Ik1}
\\
\frac{\del^2 \pi_K}{\del \Rt^2}
&=
\frac{- d}{2 (d-1)} \frac{\del \pi_K}{\del \phi(2)} \( \frac{\del \pi_K}{\del \Rt} \)^2\,.
\label{Ik2}
\ee
Meanwhile, the coefficients of the first and third line combine to give
\be
\frac{\del^2 \pi_K}{\del \phi(2)^2}
=
\frac{-d}{2 (d-1)}
\( \frac{\del \pi_K}{\del \phi(2)} \)^3
\, .
\label{Ik3}
\ee
The fourth condition (say vanishing of the second line) follows automatically from these.  To solve these three restrictions on $\pi_K$, first write~\eqref{Ik3} as
\be
\frac{\del}{\del \phi(2)} \( \frac{\del \pi_K}{\del \phi(2)} \)^{-2}
=
\frac{d}{(d-1)}
\, .
\ee
Integrating this equation yields
\be
\( \frac{\del \pi_K}{\del \phi(2)} \)^{-2}
=
\frac{d}{(d-1)} \( \phi(2) + f(t,\Rt) \)
\, ,
\ee
for some arbitrary function $f(t,\Rt)$.  This is equivalent to
\be
\frac{\del \pi_K}{\del \phi(2)}
=
\pm \sqrt{\frac{(d-1)}{d}} \frac{1}{ \sqrt{\phi(2) + f(t,\Rt)} }
\, .
\ee
Integrating this equation yields
\be
\pi_K
=
g(t, \Rt) \pm 2 \sqrt{\frac{(d-1)}{d}} \sqrt{\phi(2) + f(t,\Rt)}
\, ,
\label{pkgh}
\ee
where we have introduced a second arbitrary function, $g(t,\Rt)$.  This form for $\pi_K$ solves~\eqref{Ik3} in full generality.  Using~\eqref{pkgh}, condition~\eqref{Ik1} becomes simply
\be
\frac{\del g}{\del \Rt} = 0
\, ,
\ee
which implies that $g = g(t)$.  However, since $\pi_K$ appears in the $\Ht_i$ constraint through $\gradt_i \pi_K$, the function $g(t)$ drops out of the action~\eqref{HKaction} and can therefore be set to zero without loss of generality.  Using~\eqref{pkgh} and $g=0$, the final condition~\eqref{Ik2} becomes
\be
\frac{\del^2 f}{\del \Rt^2} = 0
\, ,
\ee
the solution to which is
\be
f(t,\Rt) = c_1(t) \Rt + c_2(t)
\, ,
\ee
where $c_1(t)$ and $c_2(t)$ are arbitrary functions of time.  The most general solution to all the restrictions on $\pi_K$ is therefore
\be
\pi_K = \pm 2 \sqrt{\frac{(d-1)}{d}}\sqrt{\phi(2) + c_1(t) \Rt + c_2(t)}
\, .
\label{pkr}
\ee
This $\pi_K$ has the same dependence on the phase space variables as the $\pi_K$ of GR --- see~\eqref{GRHK} and~\eqref{GRPW}. The only difference lies in the explicit time dependence.  The $\pi_K$ of GR depends on only one arbitrary function, $\W(t)$, while the $\pi_K$ we just derived depends on two arbitrary functions, $c_1(t)$ and $c_2(t)$. These will be further constrained in the next Section by demanding that the constraints be consistently satisfied under time evolution, leaving us with GR as the only possibility.

\section{Evolution of the Constraints}
\label{evolution}

For the constraints $\Ht_i$ to be consistent with the equations of motion, they must be preserved under time evolution, {\it i.e.},
\be
\dot{\Ht}_i \sim 0
\, .
\ee
By promoting this weak equality to a strong equation, we will derive a restriction on the allowed form of the physical Hamiltonian density $\pi_H$. 

The $\Ht_i$ generators evolve according to 
\be
\dot{\Ht}_i(x) = \frac{\del \Ht_i(x)}{\del t} + \{ \Ht_i(x) , H \}
\, ,
\label{eom}
\ee
where $H \equiv \int \ddx ( \pi_H + N^i \Ht_i )$ is the Hamiltonian.  Since the constraints are by assumption first class, $\{ \Ht_i(x) , \Ht_i(y) \}\sim 0$, and therefore
\be
\dot{\Ht}_i(x)
\sim \frac{\del \Ht_i(x)}{\del t}
+ \{ \Ht_i(x), \Pi \}
\, ,
\label{tildeHeom1}
\ee
where
\be
\Pi \equiv \int \ddx \pi_H
\, .
\ee
Because $\Pi$ coincides with $H$ when the constraints are satisfied, {\it i.e.} $\Pi \sim H$, and because it contains no constraint redundancy, $\Pi$ is called the physical Hamiltonian.
Noting that $\Ht_i$ depends explicitly on time only through $\pi_K$,~(\ref{tildeHeom1}) reduces to
\be
\dot{\Ht}_i(x)
\sim
- \del_i \( \frac{\del \pi_K(x)}{\del t} \)
+ \{ \Ht_i(x), \Pi \}
\, .
\label{hdot}
\ee
To proceed, we must calculate $\{ \Pi , \Ht_i(x) \}$. The derivation is rather technical, and we leave the details to Appendix~\ref{PiBracket}. Substituting the resulting expression~\eqref{PiH} 
in~\eqref{hdot}, we obtain 
\be
- \dot{\Ht}_i(x)
&
\sim
\del_i \(
\pi_H
+ \frac{\del \pi_K(x)}{\del t}
- \frac{2}{d} \frac{\del \pi_H}{\del \Rt} \Rt
- \sum_{n = 2}^{\infty} n \frac{\del \pi_H}{\del \phi(n)} \phi(n)
- \frac{2 (d - 1)}{d} \gradt_b \gradt^b \frac{\del \pi_H}{\del \Rt}
\right.
\nl
\qq \qq + \frac{\del \pi_H}{\del \Rt} \Rt^{j k} \sum_{m = 2}^{\infty} m \frac{\del \pi_K}{\del \phi(m)} \Pi(m-1)_{b c} \td^{b c}_{j k}
- \frac{\del \pi_K}{\del \Rt} \Rt^{b c} \sum_{n = 2}^{\infty} n \frac{\del \pi_H}{\del \phi(n)} \Pi(n-1)_{j k} \td_{b c}^{j k}
\nl \qq \qq
- \sum_{m = 2}^{\infty} m \frac{\del \pi_K}{\del \phi(m)} \Pi(m-1)_{b c} \td^{b c}_{j k} \gradt^j \gradt^k \( \frac{\del \pi_H}{\del \Rt}\)
\nl \qq \qq
\left. + \frac{\del \pi_K}{\del \Rt} \gradt^b \gradt^c \( \sum_{n = 2}^{\infty} n \frac{\del \pi_H}{\del \phi(n)} \Pi(n-1)_{j k} \td^{j k}_{b c} \) \)
.
\label{hdotbracket}
\ee
On the right hand side, everything appears beneath a single gradient operator.  The consistency condition $\dot{\Ht}_i \sim 0$ therefore implies that the quantity on which the gradient acts must 
be weakly equivalent to some function of time; call it $f(t)$.  However, we are free to redefine the quantity $\pi_H - f(t)$ to be $\pi_H$; this redefinition changes the action only by a total derivative, and leaves the equation of motion invariant.  This eliminates the arbitrary function $f(t)$, allowing us to write
\be
0
&
\sim
\pi_H
+ \frac{\del \pi_K(x)}{\del t}
- \frac{2}{d} \frac{\del \pi_H}{\del \Rt} \Rt
- \sum_{n = 2}^{\infty} n \frac{\del \pi_H}{\del \phi(n)} \phi(n)
- \frac{2 (d - 1)}{d} \gradt_b \gradt^b \frac{\del \pi_H}{\del \Rt}
\nl
+ \frac{\del \pi_H}{\del \Rt} \Rt^{j k} \sum_{m = 2}^{\infty} m \frac{\del \pi_K}{\del \phi(m)} \Pi(m-1)_{b c} \td^{b c}_{j k}
- \sum_{m = 2}^{\infty} m \frac{\del \pi_K}{\del \phi(m)} \Pi(m-1)_{b c} \td^{b c}_{j k} \gradt^j \gradt^k \( \frac{\del \pi_H}{\del \Rt}\)
\nl
- \frac{\del \pi_K}{\del \Rt} \Rt^{b c} \sum_{n = 2}^{\infty} n \frac{\del \pi_H}{\del \phi(n)} \Pi(n-1)_{j k} \td_{b c}^{j k}
+ \frac{\del \pi_K}{\del \Rt} \gradt^b \gradt^c \( \sum_{n = 2}^{\infty} n \frac{\del \pi_H}{\del \phi(n)} \Pi(n-1)_{j k} \td^{j k}_{b c} \)
.
\ee
To promote this weak equality to a strong equation, the key observation is that gradients act on loose momentum indices only in very the last term, thus all of the constraint ambiguity arises from this term.  Expanding the last term, applying the identity $\gradt_i \gradt_j M^{a b} = \gradt_j \gradt_i M^{a b} + \Rt^a_{\ c i j} M^{c b} + \Rt^b_{\ c i j} M^{a c}$, and performing the substitution $\gradt_j \pt^{i j} \gt - \half \gradt^i \pi_K$, we obtain the strong equation
\be
0
&
=
\pi_H
+ \frac{\del \pi_K(x)}{\del t}
- \frac{2}{d} \frac{\del \pi_H}{\del \Rt} \Rt
- \sum_{n = 2}^{\infty} n \frac{\del \pi_H}{\del \phi(n)} \phi(n)
- \frac{2 (d - 1)}{d} \gradt_b \gradt^b \frac{\del \pi_H}{\del \Rt}
\nl
+ \frac{1}{d} \frac{\del \pi_K}{\del \Rt} \Rt \sum_{n = 3}^{\infty} n \frac{\del \pi_H}{\del \phi(n)} \phi(n-1)
- \frac{1}{d} \frac{\del \pi_H}{\del \Rt} \Rt \sum_{m = 3}^{\infty} m \frac{\del \pi_K}{\del \phi(m)} \phi(m-1)
\nl
- \sum_{m = 2}^{\infty} m \frac{\del \pi_K}{\del \phi(m)} \Pi(m-1)_{b c} \td^{b c}_{j k} \gradt^j \gradt^k \( \frac{\del \pi_H}{\del \Rt}\)
- \frac{1}{d} \frac{\del \pi_K}{\del \Rt} \gradt_c \gradt^c \( \sum_{n = 3}^{\infty} n \frac{\del \pi_H}{\del \phi(n)} \phi(n-1) \)
\nl
- \half \frac{\del \pi_K}{\del \Rt} \( \sum_{n = 2}^{\infty} n \frac{\del \pi_H}{\del \phi(n)} \Pi(n-2)_{a b} \) \gradt^b \( \gradt^a \pi_K \)
- 2 \frac{\del \pi_K}{\del \Rt} \( \gradt^a \pi_K \) \gradt_a \( \frac{\del \pi_H}{\del \phi(2)} \)
\nl
+ \frac{\del \pi_K}{\del \Rt} (\gradt^b \pt^{a c}) \gradt_c \( \sum_{n = 3}^{\infty} n \frac{\del \pi_H}{\del \phi(n)} \Pi(n-2)_{a b} \)
+ 2 \frac{\del \pi_K}{\del \Rt} \pt^{a c} \gradt_a \gradt_c \( \frac{\del \pi_H}{\del \phi(2)} \)
\nl
+ \frac{1}{4} \frac{\del \pi_K}{\del \Rt} \( \gradt^a \pi_K \) \( \gradt^b \pi_K \) \( \sum_{n = 3}^{\infty} n \frac{\del \pi_H}{\del \phi(n)} \Pi(n-3)_{a b} \)
- \half \frac{\del \pi_K}{\del \Rt} \( \gradt^a \pi_K \) \pt^{d b} \gradt_b \( \sum_{n = 3}^{\infty} n \frac{\del \pi_H}{\del \phi(n)} \Pi(n-3)_{a d} \)
\nl
- \half \frac{\del \pi_K}{\del \Rt} \pt^{a c} \gradt_c \( \( \gradt^d \pi_K \) \( \sum_{n = 3}^{\infty} n \frac{\del \pi_H}{\del \phi(n)} \Pi(n-3)_{a d} \) \)
+ \frac{\del \pi_K}{\del \Rt} \pt^{a c} \gradt_c \( \pt^{d b} \gradt_b \( \sum_{n = 3}^{\infty} n \frac{\del \pi_H}{\del \phi(n)} \Pi(n-3)_{a d} \) \)
\nl
- 2 \frac{\del \pi_K}{\del \Rt} \frac{\del \pi_H}{\del \phi(2)} \pt_{b c} \Rt^{b c}
+ \frac{\del \pi_H}{\del \Rt} \Rt^{j k} \sum_{m = 2}^{\infty} m \frac{\del \pi_K}{\del \phi(m)} \Pi(m-1)_{j k}
+ (\pt_{c d} \Rt^{c a b d}) \frac{\del \pi_K}{\del \Rt} \( \sum_{n = 3}^{\infty} n \frac{\del \pi_H}{\del \phi(n)} \Pi(n-2)_{a b} \)
.
\label{stronghk}
\ee
We next import our knowledge of $\pi_K$ from Sec.~\ref{closure} to simplify~\eqref{stronghk}. Recall that $\pi_K$ can either take the ultra-local form $\pi_K = \pi_K(t, \phi(n))$, or the square-root form~(\ref{pkr}). We will treat each case separately. 

\subsection{Ultra-local case}

Since $\pi_K = \pi_K(t, \phi(n))$, and in particular $\del \pi_K/\del \Rt = 0$, in this case the strong equation simplifies to
\be
0
&
=
\pi_H
+ \frac{\del \pi_K}{\del t}
- \frac{2}{d} \frac{\del \pi_H}{\del \Rt} \Rt
- \sum_{n = 2}^{\infty} n \frac{\del \pi_H}{\del \phi(n)} \phi(n)
+ \frac{\del \pi_H}{\del \Rt} \Rt^{j k} \sum_{m = 2}^{\infty} m \frac{\del \pi_K}{\del \phi(m)} \Pi(m-1)_{j k}
\nl
- \frac{1}{d} \frac{\del \pi_H}{\del \Rt} \Rt \sum_{m = 3}^{\infty} m \frac{\del \pi_K}{\del \phi(m)} \phi(m-1)
- \frac{2 (d - 1)}{d} \gradt_k \gradt^k \frac{\del \pi_H}{\del \Rt}
\nl
+ \frac{1}{d} \sum_{m = 3}^{\infty} m \frac{\del \pi_K}{\del \phi(m)} \phi(m-1) \gradt_k \gradt^k \( \frac{\del \pi_H}{\del \Rt}\)
- \sum_{m = 2}^{\infty} m \frac{\del \pi_K}{\del \phi(m)} \Pi(m-1)^{j k} \gradt_j \gradt_k \( \frac{\del \pi_H}{\del \Rt}\)
.
\label{ultrastrong}
\ee
As it turns out, equation~\eqref{ultrastrong} can only be solved if $\pi_H$ is ultralocal.  To prove this, we use the same techniques involved in the derivation of equation~\eqref{pkr}.  We defer the details of the analysis to Appendix~\ref{khproof}.

Since $\pi_H$ must be ultralocal, $\del \pi_H/\del \Rt = 0$ and the strong equation~\eqref{ultrastrong} becomes
\be
0 = \pi_H + \frac{\del \pi_K}{\del t} - \sum_{n=2}^{\infty} n \frac{\del \pi_H}{\del \phi(n)} \phi(n)
\, .
\label{RGpiH}
\ee
This is the analogue of the conformal renormalization group (RG) equation discovered in~\cite{Khoury:2011ay}. It implies that the time-independent part of $\pi_K$ can be arbitrary, whereas
the time-dependent part is fixed by $\pi_H$. To summarize, the ultra-local case allows for two arbitrary ultralocal functions of the phase-space variables: a possibly time-dependent $\pi_H = \pi_H(t,\phi(n))$ and a time-independent component of $\pi_K$. This completes our analysis of the ultralocal case. 

\subsection{Local case}

We turn now to the local case, in which $\pi_K$ takes the square-root form given by~\eqref{pkr}:
\be
\pi_K = \pm 2 \sqrt{\frac{(d-1)}{d}}  \sqrt{\phi(2) + c_1(t) \Rt + c_2(t)}
\label{pikrepeat}
\, .
\ee
We of course assume that $c_1 \neq 0$, so that $\del \pi_K/\del \Rt \neq 0$, for otherwise this would reduce to the ultra-local form. Since $\pi_K$ is a function only of $t$, $\Rt$, and $\phi(2)$, the strong equation~\eqref{stronghk} can be written as
\be
0
&
=
\pi_H
+ \frac{\del \pi_K(x)}{\del t}
- \sum_{n = 2}^{\infty} n \frac{\del \pi_H}{\del \phi(n)} \phi(n)
- \frac{2}{d} \frac{\del \pi_H}{\del \Rt} \Rt
+ \frac{1}{d} \frac{\del \pi_K}{\del \Rt} \Rt \sum_{n = 3}^{\infty} n \frac{\del \pi_H}{\del \phi(n)} \phi(n-1)
- 2 \frac{\del \pi_K}{\del \Rt} \frac{\del \pi_H}{\del \phi(2)} \pt_{b c} \Rt^{b c}
\nl
+ 2 \frac{\del \pi_H}{\del \Rt} \frac{\del \pi_K}{\del \phi(2)} \pt_{j k} \Rt^{j k}
+ (\pt_{c d} \Rt^{c a b d}) \frac{\del \pi_K}{\del \Rt} \( \sum_{n = 3}^{\infty} n \frac{\del \pi_H}{\del \phi(n)} \Pi(n-2)_{a b} \)
- \frac{2 (d - 1)}{d} \gradt_k \gradt^k \frac{\del \pi_H}{\del \Rt}
\nl
- \frac{1}{d} \frac{\del \pi_K}{\del \Rt} \gradt_k \gradt^k \( \sum_{n = 3}^{\infty} n \frac{\del \pi_H}{\del \phi(n)} \phi(n-1) \)
+ 2 \frac{\del \pi_K}{\del \Rt} \pt^{j k} \gradt_j \gradt_k \( \frac{\del \pi_H}{\del \phi(2)} \)
- 2 \frac{\del \pi_K}{\del \phi(2)} \pt^{j k} \gradt_j \gradt_k \( \frac{\del \pi_H}{\del \Rt}\)
\nl
- \half \frac{\del \pi_K}{\del \Rt} \( \sum_{n = 2}^{\infty} n \frac{\del \pi_H}{\del \phi(n)} \Pi(n-2)^{a b} \) \gradt_a \gradt_b \pi_K
- \half \frac{\del \pi_K}{\del \Rt} \( \gradt^a \pi_K \) \pt^{d b} \gradt_b \( \sum_{n = 3}^{\infty} n \frac{\del \pi_H}{\del \phi(n)} \Pi(n-3)_{a d} \)
\nl
+ \frac{1}{4} \frac{\del \pi_K}{\del \Rt} \( \gradt^a \pi_K \) \( \gradt^b \pi_K \) \( \sum_{n = 3}^{\infty} n \frac{\del \pi_H}{\del \phi(n)} \Pi(n-3)_{a b} \)
- 2 \frac{\del \pi_K}{\del \Rt} \( \gradt^a \pi_K \) \gradt_a \( \frac{\del \pi_H}{\del \phi(2)} \)
\nl
- \half \frac{\del \pi_K}{\del \Rt} \pt^{a c} \gradt_c \( \( \gradt^d \pi_K \) \( \sum_{n = 3}^{\infty} n \frac{\del \pi_H}{\del \phi(n)} \Pi(n-3)_{a d} \) \)
+ 3 \frac{\del \pi_K}{\del \Rt} \pt^{a c} \gradt_c \( \pt_{a}^{\ b} \gradt_b \( \frac{\del \pi_H}{\del \phi(3)} \) \)
\nl
+ \frac{\del \pi_K}{\del \Rt} \pt^{a c} \gradt_c \( \sum_{n = 4}^{\infty} n \Pi(n-2)_{a}^{\ b} \gradt_b \frac{\del \pi_H}{\del \phi(n)} \)
+ \frac{\del \pi_K}{\del \Rt} (\gradt^b \pt^{a c}) \( \sum_{n = 3}^{\infty} n \Pi(n-2)_{a b} \gradt_c \frac{\del \pi_H}{\del \phi(n)} \)
\nl
+ \frac{\del \pi_K}{\del \Rt} (\gradt^b \pt^{a c}) \( \sum_{n = 3}^{\infty} n \frac{\del \pi_H}{\del \phi(n)} \gradt_c \Pi(n-2)_{a b} \)
+ \frac{\del \pi_K}{\del \Rt} \pt^{a c} \( \gradt_c \pt^{d b} \) \( \sum_{n = 4}^{\infty} n \frac{\del \pi_H}{\del \phi(n)} \gradt_b \Pi(n-3)_{a d} \)
\nl
+ \frac{\del \pi_K}{\del \Rt} \pt^{a c} \pt^{d b} \( \sum_{n = 4}^{\infty} n \( \gradt_c \frac{\del \pi_H}{\del \phi(n)} \) \gradt_b \Pi(n-3)_{a d} \)
+ \frac{\del \pi_K}{\del \Rt} \sum_{n = 4}^{\infty} n \frac{\del \pi_H}{\del \phi(n)} \pt^{a c} \pt^{d b} \gradt_c \gradt_b \Pi(n-3)_{a d}
\,
.
\label{localstrong}
\ee
The phase space scalars
\be
\pt^{a c} \pt^{d b} \gradt_c \gradt_b \Pi(n-3)_{a d}
\nc
n \geq 4
\ee
appear only in the final summand of~\eqref{localstrong}.  To remove these scalars from the right-hand-side of~\eqref{localstrong} without introducing new constraints on phase space, their respective coefficients
\be
\frac{\del \pi_K}{\del \Rt} n \frac{\del \pi_H}{\del \phi(n)}
\nc
n \geq 4
\ee
must vanish.  Since $\del \pi_K/\del \Rt \neq 0$, we must have
\be
\frac{\del \pi_H}{\del \phi(n)} = 0
\nc
n \geq 4
\, ,
\ee
so $\pi_H$ can depend only on $t$, $\Rt$, $\phi(2)$, and $\phi(3)$.  The strong equation~\eqref{localstrong} then becomes
\be
0
&
=
\pi_H
+ \frac{\del \pi_K(x)}{\del t}
- \sum_{n = 2}^{3} n \frac{\del \pi_H}{\del \phi(n)} \phi(n)
- \frac{2}{d} \frac{\del \pi_H}{\del \Rt} \Rt
+ 3 \frac{1}{d} \frac{\del \pi_K}{\del \Rt} \Rt \frac{\del \pi_H}{\del \phi(3)} \phi(2)
- 2 \frac{\del \pi_K}{\del \Rt} \frac{\del \pi_H}{\del \phi(2)} \pt_{b c} \Rt^{b c}
\nl
+ 2 \frac{\del \pi_H}{\del \Rt} \frac{\del \pi_K}{\del \phi(2)} \pt_{j k} \Rt^{j k}
+ 3 \( \pt_{c d} \Rt^{c a b d} \pt_{a b} \) \frac{\del \pi_K}{\del \Rt} \frac{\del \pi_H}{\del \phi(3)}
- \frac{2 (d - 1)}{d} \gradt_k \gradt^k \frac{\del \pi_H}{\del \Rt}
- 3 \frac{1}{d} \frac{\del \pi_K}{\del \Rt} \gradt_k \gradt^k \( \frac{\del \pi_H}{\del \phi(3)} \phi(2) \)
\nl
+ 2 \frac{\del \pi_K}{\del \Rt} \pt^{j k} \gradt_j \gradt_k \( \frac{\del \pi_H}{\del \phi(2)} \)
- 2 \frac{\del \pi_K}{\del \phi(2)} \pt^{j k} \gradt_j \gradt_k \( \frac{\del \pi_H}{\del \Rt}\)
- \half \frac{\del \pi_K}{\del \Rt} \( \sum_{n = 2}^{3} n \frac{\del \pi_H}{\del \phi(n)} \Pi(n-2)^{a b} \) \gradt_a \gradt_b \pi_K
\nl
- 2 \frac{\del \pi_K}{\del \Rt} \( \gradt^a \pi_K \) \gradt_a \( \frac{\del \pi_H}{\del \phi(2)} \)
+ 3 \frac{1}{4} \frac{\del \pi_K}{\del \Rt} \( \gradt^a \pi_K \) \( \gradt_a \pi_K \) \frac{\del \pi_H}{\del \phi(3)}
- 3 \half \frac{\del \pi_K}{\del \Rt} \( \gradt_a \pi_K \) \pt^{a b} \gradt_b \( \frac{\del \pi_H}{\del \phi(3)} \)
\nl
- 3 \half \frac{\del \pi_K}{\del \Rt} \pt^{a c} \gradt_c \( \( \gradt_a \pi_K \) \( \frac{\del \pi_H}{\del \phi(3)} \) \)
+ 3 \frac{\del \pi_K}{\del \Rt} \pt^{a c} \gradt_c \( \pt_{a b} \gradt^b \( \frac{\del \pi_H}{\del \phi(3)} \) \)
+ 3 \frac{\del \pi_K}{\del \Rt} \pt_{a b} (\gradt^b \pt^{a c}) \( \gradt_c \frac{\del \pi_H}{\del \phi(3)} \)
\nl
+ 3 \frac{\del \pi_K}{\del \Rt} \frac{\del \pi_H}{\del \phi(3)} (\gradt^b \pt^{a c}) ( \gradt_c \pt_{a b} )
\,
.
\label{localstrong3}
\ee
To remove the scalar $(\gradt^b \pt^{a c}) ( \gradt_c \pt_{a b} )$ from the right-hand-side of this equation without introducing new constraints on phase space, its coefficient
\be
3 \frac{\del \pi_K}{\del \Rt} \frac{\del \pi_H}{\del \phi(3)}
\ee
must vanish.  Since $\del \pi_K/\del \Rt \neq 0$, we must have
\be
\frac{\del \pi_H}{\del \phi(3)} = 0
\, ,
\ee
so $\pi_H$ can depend only on $t$, $\Rt$, and $\phi(2)$.  Using this fact, and making use of the explicit form of $\pi_K$ in all partial derivatives of $\pi_K$, the strong equation~\eqref{localstrong3} becomes
\be
0
&
=
\frac{d}{2(d-1)} \pi_K \pi_H
+ \dot{c}_1 \Rt + \dot{c}_2
- \frac{\del \pi_H}{\del \phi(2)} \phi(2) \frac{d}{(d-1)} \pi_K
- \frac{\del \pi_H}{\del \Rt} \Rt \frac{1}{(d-1)} \pi_K
- \pi_K \gradt_a \gradt^a \frac{\del \pi_H}{\del \Rt}
\nl
- c_1 \frac{\del \pi_H}{\del \phi(2)} \gradt_a \gradt^a \pi_K
- 2 c_1 \( \gradt^a \pi_K \) \( \gradt_a \frac{\del \pi_H}{\del \phi(2)} \)
+ 2 \( \frac{\del \pi_H}{\del \Rt} - c_1 \frac{\del \pi_H}{\del \phi(2)} \) \pt_{j k} \Rt^{j k}
\nl
- 2 \pt^{j k} \gradt_j \gradt_k \( \frac{\del \pi_H}{\del \Rt} - c_1 \frac{\del \pi_H}{\del \phi(2)} \)
.
\label{localstrong2}
\ee
By defining
\be
f(t,\Rt,\phi(2)) \equiv \frac{\del \pi_H}{\del \Rt} - c_1 \frac{\del \pi_H}{\del \phi(2)}
\label{fdef}
\ee
and using the chain rule, we can write the strong equation~\eqref{localstrong2} as
\be
0
&
=
\frac{d}{2(d-1)} \pi_K \pi_H
+ \dot{c}_1 \Rt + \dot{c}_2
- \frac{\del \pi_H}{\del \phi(2)} \phi(2) \frac{d}{(d-1)} \pi_K
- \frac{\del \pi_H}{\del \Rt} \Rt \frac{1}{(d-1)} \pi_K
- \pi_K \gradt_a \gradt^a \frac{\del \pi_H}{\del \Rt}
\nl
- c_1 \frac{\del \pi_H}{\del \phi(2)} \gradt_a \gradt^a \pi_K
- 2 c_1 \( \gradt^a \pi_K \) \( \gradt_a \frac{\del \pi_H}{\del \phi(2)} \)
+ 2 \pt_{j k} \Rt^{j k} f(t,\Rt,\phi(2))
- 2 \pt^{j k} \( \gradt_j \frac{\del f}{\del \Rt} \) \gradt_k \Rt
\nl
- 2 \pt^{j k} \sum_{n=2}^\infty \( \gradt_j \frac{\del f}{\del \phi(n)} \) \gradt_k \phi(n)
- 2 \frac{\del f}{\del \Rt} \pt^{j k} \gradt_j \gradt_k \Rt
- 2 \sum_{n=2}^\infty \frac{\del f}{\del \phi(n)} \pt^{j k} \gradt_j \gradt_k \phi(n)
\,
.
\label{localstrongftrp}
\ee
To eliminate the dependence of the right-hand-side on the scalars
\be
\pt^{j k} \gradt_j \gradt_k \Rt
\nc
\pt^{j k} \gradt_j \gradt_k \phi(n)
\nc
n \geq 2
\ee
without imposing any new constraints on phase space, the respective coefficients of the scalars must vanish,
\be
\frac{\del f}{\del \Rt} = 0
\nc
\frac{\del f}{\del \phi(n)} = 0
\nc
n \geq 2
\, .
\ee
It follows that $f = f(t)$, which from equation~\eqref{fdef} implies that
\be
\frac{\del \pi_H}{\del \Rt} = c_1 \frac{\del \pi_H}{\del \phi(2)} + f(t)
\, .
\label{pihrpf}
\ee
Substituting this result into the strong equation~\eqref{localstrongftrp} and simplifying a total derivative yields
\be
0
&
=
\frac{d}{2(d-1)} \pi_K \pi_H
+ \dot{c}_1 \Rt + \dot{c}_2
- \frac{\del \pi_H}{\del \phi(2)} \phi(2) \frac{d}{(d-1)} \pi_K
- \( c_1 \frac{\del \pi_H}{\del \phi(2)} + f(t) \) \Rt \frac{1}{(d-1)} \pi_K
\nl
+ 2 \pt_{j k} \Rt^{j k} f(t)
- c_1 \gradt_a \gradt^a \( \pi_K \frac{\del \pi_H}{\del \phi(2)} \)
.
\label{localstrongft}
\ee
By defining
\be
g(t,\Rt,\phi(2)) \equiv \pi_K \frac{\del \pi_H}{\del \phi(2)}
\label{gdef}
\ee
and using the chain rule, we can write the strong equation~\eqref{localstrongft} as
\be
0
&
=
\frac{d}{2(d-1)} \pi_K \pi_H
+ \dot{c}_1 \Rt + \dot{c}_2
- \frac{\del \pi_H}{\del \phi(2)} \phi(2) \frac{d}{(d-1)} \pi_K
- \( c_1 \frac{\del \pi_H}{\del \phi(2)} + f(t) \) \Rt \frac{1}{(d-1)} \pi_K
+ 2 \pt_{j k} \Rt^{j k} f(t)
\nl
- c_1 \( \gradt_a \frac{\del g}{\del \Rt} \) \gradt^a \Rt
- c_1 \sum_{n=2}^\infty \( \gradt_a \frac{\del g}{\del \phi(n)} \) \gradt^a \phi(n)
- c_1 \frac{\del g}{\del \Rt} \gradt_a \gradt^a \Rt
- c_1 \sum_{n=2}^\infty \frac{\del g}{\del \phi(n)} \gradt_a \gradt^a \phi(n)
\, .
\label{localstrongftg}
\ee
To eliminate the dependence of the right-hand-side on the scalars
\be
\gradt_a \gradt^a \Rt
\nc
\gradt_a \gradt^a \phi(n)
\nc
n \geq 2
\ee
without imposing any new constraints on phase space, the respective coefficients of the scalars must vanish,
\be
\frac{\del g}{\del \Rt} = 0
\nc
\frac{\del g}{\del \phi(n)} = 0
\nc
n \geq 2
\, .
\ee
It follows that $g = g(t)$, which from equation~\eqref{gdef} implies that
\be
\frac{\del \pi_H}{\del \phi(2)} = \frac{g(t)}{\pi_K}
\, ,
\ee
or
\be
\frac{\del \pi_H}{\del \phi(2)} = \pm g(t) \frac{1}{2} \sqrt{\frac{d}{(d-1)}} \frac{1}{\sqrt{\phi(2) + c_1(t) \Rt + c_2(t)}}
\, .
\ee
Integrating once yields
\be
\pi_H = h(t,\Rt) \pm g(t) \sqrt{\frac{d}{(d-1)}} \sqrt{\phi(2) + c_1(t) \Rt + c_2(t)}
\, ,
\ee
where $h(t,\Rt)$ is an arbitrary function.  Taking a derivative with respect to $\Rt$, we find that
\be
\frac{\del \pi_H}{\del \Rt} = \frac{\del h}{\del \Rt} \pm c_1 g(t) \half \sqrt{\frac{d}{(d-1)}} \frac{1}{ \sqrt{\phi(2) + c_1(t) \Rt + c_2(t)} }
\, ,
\ee
Comparing with equation~\eqref{pihrpf} yields
\be
\frac{\del h}{\del \Rt} = f(t)
\, ,
\ee
so
\be
h(t,\Rt) = f(t) \Rt + k(t)
\, ,
\ee
where $k(t)$ is an arbitrary function.  The physical Hamiltonian density $\pi_H$ can therefore be written as
\be
\pi_H = k(t) + f(t) \Rt \pm g(t) \sqrt{\frac{d}{(d-1)}} \sqrt{\phi(2) + c_1(t) \Rt + c_2(t)}
\ee
However, the function $k(t)$ contributes only a boundary term to the action and does not affect the equations of motion.  We can therefore choose $k = 0$ without loss of generality, in which case
\be
\pi_H = f(t) \Rt \pm g(t) \sqrt{\frac{d}{(d-1)}} \sqrt{\phi(2) + c_1(t) \Rt + c_2(t)}
\, .
\label{phf}
\ee
With this result, the strong equation~\eqref{localstrongftg} becomes
\be
0
&
=
\frac{(d-2)}{2(d-1)} \pi_K f(t) \Rt
+ g(t) c_1(t) \Rt + g(t) c_2(t) \frac{d}{(d-1)}
+ \dot{c}_1 \Rt + \dot{c}_2
+ 2 \pt_{j k} \Rt^{j k} f(t)
\, .
\label{localstrongftgt}
\ee
If $d \geq 3$, eliminating the phase space scalar $\pt_{j k} \Rt^{j k}$ from the right-hand-side of the strong equation without generating new constraints requires $f = 0$.  If $d = 2$, then $\Rt = \sqrt{\th} \Rt$ in equation~\eqref{phf} is a total derivative, and once again we can choose $f = 0$.  Since $f = 0$ either way, $\pi_H$ is
\be
\pi_H = \pm g(t) \sqrt{\frac{d}{(d-1)}} \sqrt{\phi(2) + c_1(t) \Rt + c_2(t)}
\, ,
\ee
and the strong equation becomes
\be
0
&
=
\dot{c}_2
+ c_2 \frac{d}{(d-1)} g(t)
+
\Big(
\dot{c}_1
+ c_1 g(t)
\Big)\Rt
\,
.
\label{streq2}
\ee
To eliminate the dependence of the right-hand-side of the strong equation on $\Rt$ without reducing the number of phase space degrees of freedom, its coefficient must vanish,
\be
\dot{c}_1 + g(t) c_1 = 0
\, .
\ee
Since $c_1 \neq 0$ by assumption, we conclude that:
\be
g(t) = - \frac{\dot{c}_1}{c_1}
\, .
\ee
Meanwhile, the remainder of~(\ref{streq2}) has the general solution
\be
c_2 = 2\lambda |c_1|^\frac{d}{d-1}
\, ,
\ee
where $\lambda$ is a constant.

These results are most neatly expressed in terms of a new time-dependent function:
\be
\w(t) \equiv \frac{1}{d-1} \log |c_1(t)|
\, .
\ee
Writing $c_1$ and $c_2$ in terms of $\w$ allows us to define the scalar function
\be
\pi_\w  \equiv \pm \sqrt{d (d-1)} \sqrt{\phi(2) \pm \Rt \, e^{(d-1)\w(t) } + 2 \l e^{d\w(t)}}
\, ,
\ee
in terms of which $\pi_H$ and $\pi_K$ can be written as
\be
\pi_H = - \dot{\w}(t)\pi_\w
\nc
\pi_K = \frac{2}{d} \pi_\w
\, .
\ee
Apart from the sign of the coefficient of $\Rt$ in $\pi_\W$, these last two equations are identical to~\eqref{GRHK} and~\eqref{GRPW} for GR. 
Since the $\Rt $ term determines the `gradient energy' of gravity, the sign can be fixed by requiring the absence of gradient instabilities. For instance, on looking at tensor perturbations around spatially flat FRW backgrounds, a gradient instability will only be absent for the choice of sign corresponding to GR.
The function $\w(t)$ and the parameter $\lambda$ are identified as $\W(t)$ and $\L$ of GR, respectively.

\subsection{Summary}
Demanding the consistency of the momentum constraints $\Ht_i$ with the equation of motion yields only two possibilities for $\pi_H$ and $\pi_K$.  First, $\pi_H$ and $\pi_K$ can be ultralocal functions of time $t$ and the phase space variables, subject to the restriction
\be
0 = \pi_H + \frac{\del \pi_K}{\del t} - \sum_{n=2}^{\infty} n \frac{\del \pi_H}{\del \phi(n)} \phi(n)
\, .
\label{renorm}
\ee
In this case the physical Hamiltonian does not include spatial derivatives of the fields, so there is no notion of gradient energy, and no physical coupling between adjacent spatial points.  Such theories cannot describe the local propagation of graviton degrees of freedom.

The second possibility is to have
\be
\pi_H = - \dot{\w}(t)  \pi_\w
\nc
\pi_K = \frac{2}{d}\pi_\w
\, ,
\ee
where $\w(t)$ is an arbitrary function of time, and
\be
\pi_\w \equiv \pm \sqrt{d (d-1)} \sqrt{\phi(2) \pm \Rt \, e^{(d-1)\w(t)} + 2 \lambda e^{d\w(t)}}
\, .
\ee
To leading order in local scalars, consistency alone has forced the functions $\pi_H$ and $\pi_K$ to take essentially the same form as in GR.  The local functions $\pi_H$ and $\pi_K$ obey an extended version of the restriction~\eqref{renorm}, namely
\be
0 = \pi_H + \frac{\del \pi_K}{\del t} - \sum_{n=2}^{\infty} n \frac{\del \pi_H}{\del \phi(n)} \phi(n) - \frac{2}{d} \Rt \frac{\del \pi_H}{\del \Rt}
\, .
\ee
In terms of $\pi_\w$, this becomes the RG flow equation
\be
0 = \( \frac{2}{d} \frac{\del}{\del \w} + \sum_{n=2}^{\infty} n\phi(n) \frac{\del}{\del \phi(n)}  + \frac{2}{d} \Rt \frac{\del}{\del \Rt} \) e^{- d \w/2} \pi_\w
\, .
\ee
As discussed in~\cite{Khoury:2011ay}, this equation encodes the invariance of the physical Hamiltonian density under flow through the space of conformally equivalent metrics.

\section{Conclusion}
\label{conclusion}

In this paper, we began with the spatially covariant action of the transverse, traceless graviton degrees of freedom.  We then applied the formalism developed in~\cite{Khoury:2011ay} to determine under what circumstances the momentum constraints of the theory would 1) generate a first class algebra and 2) be preserved under time evolution.  To leading order in local scalars, we found that consistency alone singles out general relativity as the unique effective field theory of the graviton degrees of freedom.

To our knowledge, this represents an enormous advance over all previous derivations of general relativity from the graviton degrees of freedom, which assume Lorentz covariance at the outset~\cite{Feynman:1996kb,Weinberg:1965rz,Deser:1969wk}.  Our approach relies on the weaker assumption of spatial covariance, and yet achieves an equally powerful result.

In light of our result, it is plausible to interpret Lorentz symmetry in the gravitational sector as an {\it accidental} or {\it emergent} symmetry.  Accidental symmetries arise in an effective field theory when all the allowable operators which violate the symmetry are confined above the cutoff of the theory; in this respect, it is the opposite of spontaneous symmetry breaking.  This is exactly the case in our derivation of general relativity, where consistency forces the operators $\phi(2)$ and $\Rt$ to appear in a Lorentz covariant combination, and our cutoff excludes all higher order operators which might spoil the symmetry.  Also of note is the fact that both the conformal scale factor $\W(t)$ and the cosmological constant $\L$ arise in the theory as constants of integration with respect to phase space scalars.

A clear direction for future research would be to generalize the derivation by allowing the functions $\pi_H$ and $\pi_K$ to depend on spatial derivatives through scalars other than $\Rt$.  There are only two logical possibilities: 1) Lorentz covariance survives at all orders in perturbation theory, or 2) Lorentz-violating scalars can modify the behavior of the graviton.  Either way, the results should be interesting.

So far, our discussion has focused solely on the graviton degrees of freedom, but any Lorentz covariant field theory could be subjected to a similar analysis.  Currently, we are investigating Hamiltonians which yield a consistent dynamical evolution for spatially covariant Yang-Mills fields~\cite{bloomfield}.  To systematically relax the assumption of spacetime symmetry throughout particle physics, it will eventually be necessary to expand our gravitational framework to include coupling between the graviton and all the relevant matter fields.\\
\\
{\bf Acknowledgments:} We thank Julian Barbour, Jolyon Bloomfield, Stanley Deser, Tim Koslowski, and C.~Jess Riedel for helpful discussions.  G.E.J.M. was supported in part by the Department of Energy under contract No. DE-AC02-76-ER-03071.  J.K. was supported in part by NASA ATP grant NNX11AI95G and the Alfred P. Sloan Foundation.  A.J.T. was supported in part by the Department of Energy under grant DE-FG02-12ER41810.

\appendix
\section{Computation of $\{ \Ht_i(x) , \Ht_a(y) \}$}
\label{algebra}

The Poisson bracket of the theory is defined to be
\be
\{ A, B \} \equiv \int \ddx \( \frac{\d A}{\d \th_{r s}(x)} \frac{\d B}{\d \pt^{r s}(x)} - \frac{\d A}{\d \pt^{r s}(x)} \frac{\d B}{\d \th_{r s}(x)}\)
\label{pb}
.
\ee
To evaluate the variational derivatives inside the Poisson bracket, one must use the relations
\be
\frac{\d \th_{i j}(x)}{\d \th_{r s}(z)} = \td^{r s}_{i j} \d^d(x-z)
\nc
\frac{\d \pt^{i j}(x)}{\d \pt^{r s}(z)} = \td^{i j}_{r s} \d^d(x-z)
\, ,
\ee
and
\be
\frac{\d \pt^{i j}(x)}{\d \th_{r s}(z)} = -\frac{1}{d} \th^{i j} \pt^{r s} \d^d(x-z)
\nc
\frac{\d \th_{i j}(x)}{\d \pt^{r s}(z)} = 0
\, ,
\ee
from which follow the operator identities
\be
\frac{\d}{\d \th_{i j}} = \td_{a b}^{i j} \frac{\d}{\d \th_{a b}} \, , \qq \qq
\frac{\d}{\d \pt^{i j}} = \td^{a b}_{i j} \frac{\d}{\d \pt^{a b}} \, .
\ee
The symbol $\td^{i j}_{a b}$ denotes the traceless, symmetric Kronecker matrix,
\be
\td^{i j}_{a b} = \d^{i j}_{a b} - \frac{1}{d} \th^{i j} \th_{a b}
\, ,
\ee
while $\d^{i j}_{a b}$ denotes the symmetric Kronecker matrix,
\be
\d^{i j}_{a b} = \half \( \d^i_a \d^j_b + \d^j_a \d^i_b \)
.
\ee
To compute the Poisson bracket $\{ \Ht_i(x) , \Ht_a(y) \}$, we first use the functionals
\be
F \equiv \int \ddx f^i \Ht_i
\nc
G \equiv \int \ddy g^a \Ht_a
\, ,
\ee
to compute
\be
\{ F , G \} = \int \ddx \ddy f^i(x) g^a(y) \{ \Ht_i(x) , \Ht_a(y) \}
\, .
\ee
Write the momentum constraints in the action~\eqref{HKaction} as
\be
\Ht_i = \J_i + \K_i
\nc
\J_i \equiv- 2 \th_{i j} \gradt_{k} \pt^{j k}
\nc
\K_i \equiv - \gradt_i \pi_K
\, .
\label{hjk}
\ee
This allows us to write $F = F_J + F_K$ and $G = G_J + G_K$, where
\be
F_J \equiv \int \ddx f^i \J_i
\nc
F_K \equiv \int \ddx f^i \K_i
\, ,
\ee
and
\be
G_J \equiv \int \ddy g^a \J_a
\nc
G_K \equiv \int \ddy g^a \K_a
\, .
\ee
It follows that
\be
\{ F , G \} = \{ F_J , G_J \} + \{ F_J , G_K \} + \{ F_K , G_J \} + \{ F_K , G_K \}
\, .
\ee
Recall that $\J_i \equiv -2 \th_{i j} \gradt_k \pt^{j k}$.  It is straightforward to verify that
\be
\frac{\d F_J}{\d \th_{r s}}
&
= \td^{r s}_{i j} \( 2 \pt^{j k} \gradt_k f^i \) - \gradt_i \( f^i \pt^{r s} \) - \frac{2}{d} \pt^{r s} \gradt_i f^i
\, , \nn \\
\frac{\d F_J}{\d \pt^{r s}}
&
= \td^{j k}_{r s} \(2 \th_{i j} \gradt_k f^i \)
,
\label{varfj}
\ee
and
\be
\frac{\d G_J}{\d \th_{r s}}
&
= \td^{r s}_{a b} \( 2 \pt^{b c} \gradt_c g^a \) - \gradt_a \( g^a \pt^{r s} \) - \frac{2}{d} \pt^{r s} \gradt_a g^a
\, , \nn \\
\frac{\d G_J}{\d \pt^{r s}}
&
= \td^{b c}_{r s} \(2 \th_{a b} \gradt_c g^a \)
.
\label{vargj}
\ee
Recall that $\K_i \equiv - \gradt_i \pi_K$.  If $\pi_K = \pi_K(t,\phi(n),\Rt)$, then
\be
\frac{\d F_K}{\d \th_{r s}}
&
= (\del_i f^i) \sum_{n = 2}^{\infty} n \frac{\del \pi_K}{\del \phi(n)} \( \td^{r s}_{j k} \Pi(n)^{j k} - \frac{1}{d} \pt^{r s} \phi(n-1) \)
\nl
- (\del_i f^i) \frac{\del \pi_K}{\del \Rt} \td_{j k}^{r s} \Rt^{j k}
+ \td_{j k}^{r s} \gradt^j \gradt^k \( (\del_i f^i) \frac{\del \pi_K}{\del \Rt}\)
\nn \\
\frac{\d F_K}{\d \pt^{r s}}
&
= (\del_i f^i) \sum_{n = 2}^{\infty} n \frac{\del \pi_K}{\del \phi(n)} \td^{j k}_{r s} \Pi(n-1)_{j k}
\, ,
\ee
and
\be
\frac{\d G_K}{\d \th_{r s}}
&
= (\del_a g^a) \sum_{m = 2}^{\infty} m \frac{\del \pi_K}{\del \phi(m)} \( \td^{r s}_{b c} \Pi(m)^{b c} - \frac{1}{d} \pt^{r s} \phi(m-1) \)
\nl
- (\del_a g^a) \frac{\del \pi_K}{\del \Rt} \td_{b c}^{r s} \Rt^{b c}
+ \td_{b c}^{r s} \gradt^b \gradt^c \( (\del_a g^a) \frac{\del \pi_K}{\del \Rt}\)
\nn \\
\frac{\d G_K}{\d \pt^{r s}}
&
= (\del_a g^a) \sum_{m = 2}^{\infty} m \frac{\del \pi_K}{\del \phi(m)} \td^{b c}_{r s} \Pi(m-1)_{b c}
\, .
\label{vargk}
\ee
Combining the $F_J$ and $G_J$ variations into the bracket $\{ F_J, G_J \}$ yields
\be
\{ F_J, G_J \} = 2 \int \ddz & \left\{ \(\gradt_c f^i \) \(\gradt_i g^a \) \th_{a b} \pt^{b c} - \(\gradt_k g^a \) \(\gradt_a f^i \) \th_{i j} \pt^{j k} \right. \nn \\
 &\left.+ \(\gradt_k f^i\) \gradt_a \( g^a \th_{i j} \pt^{j k} \) - \(\gradt_c g^a\) \gradt_i \( f^i \th_{a b} \pt^{b c} \) \right\} .
\ee
After integrating by parts, using the definition $\J_i = - 2 \th_{i j} \gradt_k \pt^{j k}$, and using the identity $\(\gradt_i \gradt_j - \gradt_j \gradt_i \) V^a = \Rt^a_{\ b i j} V^b$, this reduces to
\be
\{ F_J, G_J \} = &\int \ddz \left\{ f^i \J_a \gradt_i g^a - g^a \J_i \gradt_a f^i + 2 f^i g^a \pt^{j k} \( \Rt_{j i k a} + \Rt_{j a i k} \) \right\} .
\ee
From the symmetries of the Riemann tensor\footnote{$\Rt_{a b c d} = \Rt_{c d a b}$, $\Rt_{a b c d} = - \Rt_{b a c d} = - \Rt_{a b d c}$.} and the traceless momentum tensor\footnote{$\pt^{i j} = \pt^{j i}$.}, it follows that $\pt^{j k} \( \Rt_{j i k a} + \Rt_{j a i k} \) = 0$, so the last term in the integrand vanishes.  The connection terms inside the remaining covariant derivatives cancel to yield
\be
\{ F_J, G_J \} = \int \ddz f^i \J_a \del_i g^a
- \int \ddz g^a \J_i \del_a f^i
\, .
\ee
To extract the bracket $\{ \J_{i}(x), \J_{a}(y) \}$ from this result, first relabel dummy indices
\be
\{ F_J, G_J \} = \int \dcx f^i \J_a \del_i g^a - \int \dcy g^a \J_i \del_a f^i \, .
\ee
Under the spatial derivatives in this equation, insert the identities
\be
g^a(x) = \int \dcy \d^3(x-y) g^a(y) \, , \qquad f^i(y) = \int \dcx \d^3(x-y) f^i(x) \, , \label{deltaeq}
\ee
to obtain
\be
\{ F_J, G_J \} = \int \dcx \dcy f^i(x) g^a(y) \( \J_a(x) \del_{x^i} \d^3(x-y) - \J_i(y) \del_{y^a} \d^3(x-y) \) .
\ee
This yields the distributional identity
\be
\{ \J_i(x),\J_a(y) \} = \J_a(x) \del_{x^i} \d^3(x-y) - \J_i(y) \del_{y^a} \d^3(x-y) \, . \label{jj}
\ee
Combining the $F_J$ and $G_K$ variations into the bracket $\{ F_J, G_K \}$ yields
\be
\{ F_J , G_K \}
&
=
- \int \ddz (\del_i f^i) (\del_a g^a) \sum_{n = 2}^{\infty} n \frac{\del \pi_K}{\del \phi(n)} \phi(n)
- \frac{2}{d} \int \ddz ( \del_i f^i ) (\del_a g^a) \frac{\del \pi_K}{\del \Rt} \Rt
\nl
- \int \ddz f^i (\del_a g^a) \sum_{n = 2}^{\infty} \frac{\del \pi_K}{\del \phi(n)} \del_i \phi(n)
+ 2 \int \ddz (\del_a g^a) ( \Rt_{i k} \gradt^k f^i ) \frac{\del \pi_K}{\del \Rt}
\nl
- 2 \int \ddz \( \gradt_k f^i \) \gradt_i \gradt^k \( (\del_a g^a) \frac{\del \pi_K}{\del \Rt}\)
+ \frac{2}{d} \int \ddz ( \del_i f^i ) \gradt_c \gradt^c \( (\del_a g^a) \frac{\del \pi_K}{\del \Rt}\)
.
\ee
Using the identities $\gradt_i \gradt_k f^i = \gradt_k \gradt_i f^i + \Rt_{i k} f^i$ and $2 \gradt^k \Rt_{i k} = \gradt_i \Rt$, integrating by parts, simplifying a total derivative of $\pi_K$, integrating by parts again, and expanding yields
\be
\{ F_J , G_K \}
&
=
- \int \ddz f^i (\del_a g^a) \del_i \pi_K
+ \frac{2 (d-1)}{d} \int \ddz (\del_a g^a) \( \gradt_k (\del_i f^i ) \) \gradt^k \frac{\del \pi_K}{\del \Rt}
\nl
+ \frac{2 (d-1)}{d} \int \ddz \frac{\del \pi_K}{\del \Rt} \( \gradt_k (\del_i f^i ) \) \( \gradt^k (\del_a g^a) \)
\nl
- \int \ddz ( \del_i f^i ) (\del_a g^a) \( \frac{2}{d} \frac{\del \pi_K}{\del \Rt} \Rt + \sum_{n = 2}^{\infty} n \frac{\del \pi_K}{\del \phi(n)} \phi(n) \) ,
\ee
and a parallel expression for $\{ F_K , G_J \}$.  The sum of the two brackets reduces to
\be
\{ F_J , G_K \} + \{ F_K , G_J \}
&
=
- \int \ddz (\del_a g^a) f^i \del_i \pi_K
+
\int \ddz (\del_i f^i) g^a \del_a \pi_K
\nl
+ \frac{2 (d-1)}{d} \int \ddz (\del_a g^a) \( \gradt_k (\del_i f^i ) \) \gradt^k \frac{\del \pi_K}{\del \Rt}
\nl
- \frac{2 (d-1)}{d} \int \ddz (\del_i f^i) \( \gradt_k (\del_a g^a ) \) \gradt^k \frac{\del \pi_K}{\del \Rt}
\, .
\ee
After integrating by parts, expanding, and simplifying, we obtain
\be
\{ F_J , G_K \} + \{ F_K , G_J \}
&
=
\int \ddz f^i \K_a \del_i g^a
- \int \ddz g^a \K_i \del_a f^i
\nl
+ \int \ddz (\del_i f^i) (\del_k \del_a g^a) \M^k
- \int \ddz (\del_a g^a) (\del_k \del_i f^i) \M^k
\, ,
\ee
where
\be
\M_k \equiv - \frac{2 (d-1)}{d} \gradt_k \frac{\del \pi_K}{\del \Rt}
\, .
\ee
To extract the bracket $\{ \J_i(x), \K_a(y) \}+\{ \K_i(x), \J_a(y) \}$, integrate by parts, relabel dummy indices, and insert the identities in equation~(\ref{deltaeq}) to yield
\be
\{ F_J , G_K \} + \{ F_K , G_J \}
&
= \int \dcx \dcy f^i(x) g^a(y) \( \K_a(x) \del_{x^i} \d^3(x-y) - \K_i(y) \del_{y^a} \d^3(x-y) \)
\nl
+\int \dcx \dcy f^i(x) g^a(y) \del_{x^i} \( - \M^k(x) \del_{x^k} \del_{x^a} \d^3(x-y) \)
\nl
- \int \dcx \dcy f^i(x) g^a(y) \del_{y^a} \( - \M^k(y) \del_{y^k} \del_{y^i} \d^3(x-y) \) .
\ee
It is clear that
\be
\{ \J_i(x),\K_a(y) \} + \{ \K_i(x),\J_a(y) \}
&
= \K_a(x) \del_{x^i} \d^3(x-y) - \K_i(y) \del_{y^a} \d^3(x-y)
\nl
+ \del_{x^i} \( - \M^k(x) \del_{x^k} \del_{x^a} \d^3(x-y) \)
\nl
- \del_{y^a} \( - \M^k(y) \del_{y^k} \del_{y^i} \d^3(x-y) \)
.
\label{rjk}
\ee
Combining the $F_K$ and $G_K$ variations into the bracket $\{ F_K, G_K \}$ yields
\be
\{ F_K , G_K \}
&
=
\int \ddz (\del_a g^a) \( \sum_{m = 2}^{\infty} m \frac{\del \pi_K}{\del \phi(m)} \Pi(m-1)_{b c} \td_{j k}^{b c} \) \gradt^j \gradt^k \( (\del_i f^i) \frac{\del \pi_K}{\del \Rt}\)
\nl
- \int \ddz (\del_i f^i) \( \sum_{m = 2}^{\infty} m \frac{\del \pi_K}{\del \phi(m)} \Pi(m-1)_{b c} \td_{j k}^{b c} \) \gradt^j \gradt^k \( (\del_a g^a) \frac{\del \pi_K}{\del \Rt}\)
\ee
Integrating by parts, expanding, and simplifying yields
\be
\{ F_K , G_K \}
&
=
\int \ddz (\gradt^j \del_i f^i) (\del_a g^a) \( \sum_{m = 2}^{\infty} m \frac{\del \pi_K}{\del \phi(m)} \Pi(m-1)_{b c} \td_{j k}^{b c} \)    \( \gradt^k \frac{\del \pi_K}{\del \Rt}\)
\nl
- \int \ddz (\gradt^k \del_i f^i) (\del_a g^a) \( \frac{\del \pi_K}{\del \Rt} \gradt^j \( \sum_{m = 2}^{\infty} m \frac{\del \pi_K}{\del \phi(m)} \Pi(m-1)_{b c} \td_{j k}^{b c} \) \)
\nl
+
\int \ddz (\del_i f^i) (\gradt^k \del_a g^a) \( \frac{\del \pi_K}{\del \Rt} \gradt^j \( \sum_{m = 2}^{\infty} m \frac{\del \pi_K}{\del \phi(m)} \Pi(m-1)_{b c} \td_{j k}^{b c} \)  \) 
\nl
- \int \ddz (\del_i f^i) (\gradt^j \del_a g^a) \( \sum_{m = 2}^{\infty} m \frac{\del \pi_K}{\del \phi(m)} \Pi(m-1)_{b c} \td_{j k}^{b c} \) \( \gradt^k \frac{\del \pi_K}{\del \Rt}\)
\ee
Define
\be
\N_k
&
\equiv
\frac{\del \pi_K}{\del \Rt} \gradt^j \( \sum_{m = 2}^{\infty} m \frac{\del \pi_K}{\del \phi(m)} \Pi(m-1)_{b c} \td_{j k}^{b c} \)
\nl
- \( \sum_{m = 2}^{\infty} m \frac{\del \pi_K}{\del \phi(m)} \Pi(m-1)_{b c} \td_{j k}^{b c} \) \gradt^j \frac{\del \pi_K}{\del \Rt}
\ee
Then
\be
\{ F_K , G_K \}
&
=
\int \ddz (\del_i f^i) (\del_k \del_a g^a) \N^k
- \int \ddz (\del_k \del_i f^i) (\del_a g^a) \N^k
\ee
To extract the bracket $\{ \K_i(x), \K_a(y) \}$, integrate by parts, relabel dummy indices, and insert the identities in equation~(\ref{deltaeq}) to obtain
\be
\{ F_K , G_K \}
&
= \int \dcx \dcy f^i(x) g^a(y) \del_{x^i} \( - \N^k(x) \del_{x^k} \del_{x^a} \d^3(x-y) \)
\nl
- \int \dcx \dcy f^i(x) g^a(y) \del_{y^a} \( - \N^k(y) \del_{y^k} \del_{y^i} \d^3(x-y) \) .
\ee
It follows that
\be
\{ \K_i(x), \K_a(y) \}
&
= \del_{x^i} \( - \N^k(x) \del_{x^k} \del_{x^a} \d^3(x-y) \)
\nl
- \del_{y^a} \( - \N^k(y) \del_{y^k} \del_{y^i} \d^3(x-y) \) \, . \label{rkk}
\ee
Since $\Ht_i = \J_i + \K_i$, combining equation~\eqref{jj} with equations~\eqref{rjk} and~\eqref{rkk} yields
\be
\{ \Ht_i(x) , \Ht_a(y) \}
&
= \Ht_a(x) \del_{x^i} \d^3(x-y) - \Ht_i(y) \del_{y^a} \d^3(x-y)
\nl
+ \del_{x^i} \( - \I^k(x) \del_{x^k} \del_{x^a} \d^3(x-y) \)
\nl
- \del_{y^a} \( - \I^k(y) \del_{y^k} \del_{y^i} \d^3(x-y) \)
,
\label{HtiHta}
\ee
where $\I_k \equiv \M_k + \N_k$.  In terms of $\pi_K$,
\be
\I_k
&
=
\frac{\del \pi_K}{\del \Rt} \gradt^j \( \sum_{m = 2}^{\infty} m \frac{\del \pi_K}{\del \phi(m)} \Pi(m-1)_{b c} \td_{j k}^{b c} \)
\nl
- \( \sum_{m = 2}^{\infty} m \frac{\del \pi_K}{\del \phi(m)} \Pi(m-1)_{b c} \td_{j k}^{b c} \) \gradt^j \frac{\del \pi_K}{\del \Rt}
- \frac{2 (d-1)}{d} \gradt_k \frac{\del \pi_K}{\del \Rt}
\, .
\label{Ik}
\ee

\section{Computation of $\{ \Pi , \Ht_i(x) \}$}
\label{PiBracket}

Recall that the physical Hamiltonian is
\be
\Pi = \int \ddx \pi_H
\, .
\ee
If $\pi_H = \pi_H(t, \phi(n),\Rt)$, the variational derivatives of $\Pi$ are
\be
\frac{\d \Pi}{\d \th_{r s}}
&
= \sum_{n = 2}^{\infty} n \frac{\del \pi_H}{\del \phi(n)} \( \td^{r s}_{j k} \Pi(n)^{j k} - \frac{1}{d} \pt^{r s} \phi(n-1) \)
- \frac{\del \pi_H}{\del \Rt} \td_{j k}^{r s} \Rt^{j k}
+ \td_{j k}^{r s} \gradt^j \gradt^k \( \frac{\del \pi_H}{\del \Rt}\)
\nn \\
\frac{\d \Pi}{\d \pt^{r s}}
&
= \sum_{n = 2}^{\infty} n \frac{\del \pi_H}{\del \phi(n)} \td^{j k}_{r s} \Pi(n-1)_{j k}
\, .
\label{varPi}
\ee
Recall from equation~\eqref{hjk} that
\be
\Ht_i = \J_i + \K_i
\nc
\J_i \equiv- 2 \th_{i j} \gradt_{k} \pt^{j k}
\nc
\K_i \equiv - \gradt_i \pi_K
\, .
\ee
To compute $\{ \Pi , \Ht_i(x) \}$, we first compute the brackets $\{ \Pi , \J_i(x) \}$ and $\{ \Pi , \K_i(x) \}$.  Combining equations~\eqref{varPi} and~\eqref{vargj} yields the Poisson bracket
\be
\{ \Pi , G_J \}
&
=
\int \ddz g^a \sum_{n = 2}^{\infty} \frac{\del \pi_H}{\del \phi(n)} \gradt_a \phi(n)
+ \int \ddz \( \gradt_a g^a \) \(
\frac{2}{d} \frac{\del \pi_H}{\del \Rt} \Rt
+ \sum_{n = 2}^{\infty} n \frac{\del \pi_H}{\del \phi(n)} \phi(n) \)
\nl
- \int \ddz 2 \( \gradt^b g^a \) \frac{\del \pi_H}{\del \Rt} \Rt_{a b}
- \int \ddz \frac{2}{d} \( \gradt_a g^a \) \gradt_k \gradt^k \( \frac{\del \pi_H}{\del \Rt}\)
\nl
+ \int \ddz 2 \( \gradt_b g^a \) \gradt_a \gradt^b \( \frac{\del \pi_H}{\del \Rt}\)
.
\ee
Integrating by parts, using the identity $\gradt_a \gradt_b g^a = \gradt_b \gradt_a g^a + g^a \Rt_{a b}$, and integrating by parts again yields
\be
\{ \Pi , G_J \}
&
=
\int \ddz g^a \sum_{n = 2}^{\infty} \frac{\del \pi_H}{\del \phi(n)} \gradt_a \phi(n)
+ \int \ddz \( \gradt_a g^a \) \(
\frac{2}{d} \frac{\del \pi_H}{\del \Rt} \Rt
+ \sum_{n = 2}^{\infty} n \frac{\del \pi_H}{\del \phi(n)} \phi(n) \)
\nl
+ \int \ddz 2 g^a \frac{\del \pi_H}{\del \Rt} \gradt^b \Rt_{a b}
+ \frac{2 (d - 1)}{d} \int \ddz \( \gradt_a g^a \) \gradt_b \gradt^b \( \frac{\del \pi_H}{\del \Rt}\)
.
\ee
Since $2 \gradt^b \Rt_{a b} = \gradt_a \Rt = \del_a \Rt$ and $\gradt_a \phi(n) = \del_a \phi(n)$, 
\be
\{ \Pi , G_J \}
&
=
\int \ddz g^a \(
\frac{\del \pi_H}{\del \Rt} \del_a \Rt
+
\sum_{n = 2}^{\infty} \frac{\del \pi_H}{\del \phi(n)} \del_a \phi(n)
\)
\nl
+ \int \ddz \( \gradt_a g^a \) \(
\frac{2}{d} \frac{\del \pi_H}{\del \Rt} \Rt
+ \sum_{n = 2}^{\infty} n \frac{\del \pi_H}{\del \phi(n)} \phi(n) \)
\nl
+ \frac{2 (d - 1)}{d} \int \ddz \( \gradt_a g^a \) \gradt_b \gradt^b \( \frac{\del \pi_H}{\del \Rt}\)
.
\ee
Simplifying a total derivative and integrating by parts yields
\be
\{ \Pi , G_J \}
&
=
\int \ddz g^a \del_a \pi_H
- \int \ddz g^a \del_a \(
\frac{2}{d} \frac{\del \pi_H}{\del \Rt} \Rt
+ \sum_{n = 2}^{\infty} n \frac{\del \pi_H}{\del \phi(n)} \phi(n) \)
\nl
- \int \ddz g^a \del_a \( \frac{2 (d - 1)}{d} \gradt_b \gradt^b \frac{\del \pi_H}{\del \Rt}\)
.
\ee
It follows that
\be
\{ \Pi , \J_i \}
=
\del_i \(
\pi_H
- \frac{2}{d} \frac{\del \pi_H}{\del \Rt} \Rt
- \sum_{n = 2}^{\infty} n \frac{\del \pi_H}{\del \phi(n)} \phi(n)
- \frac{2 (d - 1)}{d} \gradt_b \gradt^b \frac{\del \pi_H}{\del \Rt}\)
.
\label{PiJ}
\ee
Combining equations~\eqref{varPi} and~\eqref{vargk} yields the Poisson bracket
\be
\{ \Pi , G_K \}
&
= - \int \ddz (\del_a g^a) \( \sum_{m = 2}^{\infty} m \frac{\del \pi_K}{\del \phi(m)} \Pi(m-1)_{b c} \td^{b c}_{j k} \) \frac{\del \pi_H}{\del \Rt} \Rt^{j k}
\nl%
+ \int \ddz (\del_a g^a) \( \sum_{m = 2}^{\infty} m \frac{\del \pi_K}{\del \phi(m)} \Pi(m-1)_{b c} \td^{b c}_{j k} \) \gradt^j \gradt^k \( \frac{\del \pi_H}{\del \Rt}\)
\nl%
+ \int \ddz (\del_a g^a) \( \sum_{n = 2}^{\infty} n \frac{\del \pi_H}{\del \phi(n)} \Pi(n-1)_{j k} \td_{b c}^{j k} \) \frac{\del \pi_K}{\del \Rt} \Rt^{b c}
\nl%
- \int \ddz \( \sum_{n = 2}^{\infty} n \frac{\del \pi_H}{\del \phi(n)} \Pi(n-1)_{j k} \td^{j k}_{b c} \) \gradt^b \gradt^c \( (\del_a g^a) \frac{\del \pi_K}{\del \Rt}\)
\ee
Repeated integration by parts yields
\be
\{ \Pi , G_K \}
&
= \int \ddz g^a \del_a \( \frac{\del \pi_H}{\del \Rt} \Rt^{j k} \sum_{m = 2}^{\infty} m \frac{\del \pi_K}{\del \phi(m)} \Pi(m-1)_{b c} \td^{b c}_{j k} \)
\nl%
- \int \ddz g^a \del_a \( \frac{\del \pi_K}{\del \Rt} \Rt^{b c} \sum_{n = 2}^{\infty} n \frac{\del \pi_H}{\del \phi(n)} \Pi(n-1)_{j k} \td_{b c}^{j k} \)
\nl%
- \int \ddz g^a \del_a \( \sum_{m = 2}^{\infty} m \frac{\del \pi_K}{\del \phi(m)} \Pi(m-1)_{b c} \td^{b c}_{j k} \gradt^j \gradt^k \( \frac{\del \pi_H}{\del \Rt}\) \)
\nl%
+ \int \ddz g^a \del_a \( \frac{\del \pi_K}{\del \Rt} \gradt^b \gradt^c \( \sum_{n = 2}^{\infty} n \frac{\del \pi_H}{\del \phi(n)} \Pi(n-1)_{j k} \td^{j k}_{b c} \) \)
\ee
It follows that
\be
\{ \Pi , \K_i \}
&
= \del_i \( \frac{\del \pi_H}{\del \Rt} \Rt^{j k} \sum_{m = 2}^{\infty} m \frac{\del \pi_K}{\del \phi(m)} \Pi(m-1)_{b c} \td^{b c}_{j k}
- \frac{\del \pi_K}{\del \Rt} \Rt^{b c} \sum_{n = 2}^{\infty} n \frac{\del \pi_H}{\del \phi(n)} \Pi(n-1)_{j k} \td_{b c}^{j k}
\right.
\nl \qq
- \sum_{m = 2}^{\infty} m \frac{\del \pi_K}{\del \phi(m)} \Pi(m-1)_{b c} \td^{b c}_{j k} \gradt^j \gradt^k \( \frac{\del \pi_H}{\del \Rt}\)
\nl \qq
\left. + \frac{\del \pi_K}{\del \Rt} \gradt^b \gradt^c \( \sum_{n = 2}^{\infty} n \frac{\del \pi_H}{\del \phi(n)} \Pi(n-1)_{j k} \td^{j k}_{b c} \) \)
.
\label{PiK}
\ee
Combining equations~\eqref{PiJ} and~\eqref{PiK} yields
\be
\{ \Pi , \Ht_i \}
&
=
\del_i \(
\pi_H
- \frac{2}{d} \frac{\del \pi_H}{\del \Rt} \Rt
- \sum_{n = 2}^{\infty} n \frac{\del \pi_H}{\del \phi(n)} \phi(n)
- \frac{2 (d - 1)}{d} \gradt_b \gradt^b \frac{\del \pi_H}{\del \Rt} \)
\nl
+
\del_i \( \frac{\del \pi_H}{\del \Rt} \Rt^{j k} \sum_{m = 2}^{\infty} m \frac{\del \pi_K}{\del \phi(m)} \Pi(m-1)_{b c} \td^{b c}_{j k}
- \frac{\del \pi_K}{\del \Rt} \Rt^{b c} \sum_{n = 2}^{\infty} n \frac{\del \pi_H}{\del \phi(n)} \Pi(n-1)_{j k} \td_{b c}^{j k}
\right.
\nl \qq \qq
- \sum_{m = 2}^{\infty} m \frac{\del \pi_K}{\del \phi(m)} \Pi(m-1)_{b c} \td^{b c}_{j k} \gradt^j \gradt^k \( \frac{\del \pi_H}{\del \Rt}\)
\nl \qq \qq
\left. + \frac{\del \pi_K}{\del \Rt} \gradt^b \gradt^c \( \sum_{n = 2}^{\infty} n \frac{\del \pi_H}{\del \phi(n)} \Pi(n-1)_{j k} \td^{j k}_{b c} \) \)
\, .
\label{PiH}
\ee

\section{Proof that ultralocal $\pi_K$ requires ultralocal $\pi_H$}
\label{khproof}

In this section, we prove that if $\pi_K$ is ultralocal, then $\pi_H$ must be ultralocal.  When $\pi_K$ is ultralocal, the two functions $\pi_K$ and $\pi_H$ must satisfy equation~\eqref{ultrastrong}, namely
\be
0
&
=
\pi_H
+ \frac{\del \pi_K}{\del t}
- \frac{2}{d} \frac{\del \pi_H}{\del \Rt} \Rt
- \sum_{n = 2}^{\infty} n \frac{\del \pi_H}{\del \phi(n)} \phi(n)
+ \frac{\del \pi_H}{\del \Rt} \Rt^{j k} \sum_{m = 2}^{\infty} m \frac{\del \pi_K}{\del \phi(m)} \Pi(m-1)_{j k}
\nl
- \frac{1}{d} \frac{\del \pi_H}{\del \Rt} \Rt \sum_{m = 3}^{\infty} m \frac{\del \pi_K}{\del \phi(m)} \phi(m-1)
- \frac{2 (d - 1)}{d} \gradt_k \gradt^k \frac{\del \pi_H}{\del \Rt}
\nl
+ \frac{1}{d} \sum_{m = 3}^{\infty} m \frac{\del \pi_K}{\del \phi(m)} \phi(m-1) \gradt_k \gradt^k \( \frac{\del \pi_H}{\del \Rt}\)
- \sum_{m = 2}^{\infty} m \frac{\del \pi_K}{\del \phi(m)} \Pi(m-1)^{j k} \gradt_j \gradt_k \( \frac{\del \pi_H}{\del \Rt}\)
.
\label{proofstrong}
\ee
The right-hand-side of this equation depends on an infinite number of distinct phase space scalars, but the left-hand-side is $0$, which does not depend on any phase space scalars.  It follows that the right-hand-side cannot depend on any phase space scalars either.  We cannot impose additional restrictions on the phase space scalars themselves without reducing the number of physical degrees of freedom, so eliminating dependence of the right-hand-side on a particular phase space scalar can only impose restrictions on $\pi_K$ or $\pi_H$.  Instead of trying to derive all the restrictions at once, we will focus on eliminating the dependence of~\eqref{proofstrong} on a few types of scalars at a time.  Using the chain rule, the final summand of~\eqref{proofstrong} can be written as
\be
\sum_{m = 2}^{\infty} m \frac{\del \pi_K}{\del \phi(m)} \Pi(m-1)^{j k} \gradt_j \gradt_k \( \frac{\del \pi_H}{\del \Rt}\)
&
= \sum_{m = 2}^{\infty} m \frac{\del \pi_K}{\del \phi(m)} \frac{\del^2 \pi_H}{\del \Rt^2} \Pi(m-1)^{j k} \gradt_j \gradt_k \Rt
\nl
+ \sum_{m = 2}^{\infty} \sum_{n=2}^\infty m \frac{\del \pi_K}{\del \phi(m)} \frac{\del^2 \pi_H}{\del \phi(n) \del \Rt} \Pi(m-1)^{j k} \gradt_j \gradt_k \phi(n)
\nl
+ \sum_{m = 2}^{\infty} m \frac{\del \pi_K}{\del \phi(m)} \Pi(m-1)^{j k} \( \gradt_j \frac{\del^2 \pi_H}{\del \Rt^2} \) \gradt_k \Rt
\nl
+ \sum_{m = 2}^{\infty} m \frac{\del \pi_K}{\del \phi(m)} \Pi(m-1)^{j k} \sum_{n=2}^\infty \( \gradt_j \frac{\del^2 \pi_H}{\del \phi(n) \del \Rt} \) \gradt_k \phi(n)
\, .
\ee
By inspection, the scalars
\be
\Pi(m-1)^{j k} \gradt_j \gradt_k \Rt
\nc
\Pi(m-1)^{j k} \gradt_j \gradt_k \phi(n)
\nc
m,n \geq 2
\ee
appear in the strong equation~\eqref{proofstrong} only inside this final summand, with scalar coefficients
\be
m \frac{\del \pi_K}{\del \phi(m)} \frac{\del^2 \pi_H}{\del \Rt^2}
\nc
m \frac{\del \pi_K}{\del \phi(m)} \frac{\del^2 \pi_H}{\del \phi(n) \del \Rt}
\, .
\label{strongco}
\ee
To satisfy the strong equation~\eqref{proofstrong}, these coefficients must vanish.  To prove that $\pi_H$ must be ultralocal, we will examine two mutually exclusive and exhaustive cases: first, the case when $\del \pi_K/\del \phi(n) = 0$ for all $n$; second, the case when $\del \pi_K/\del \phi(n) \neq 0$ for some $n$.

\subsection{$\del \pi_K/\del \phi(n) = 0$ for all $n$}
One way the coefficients~\eqref{strongco} can vanish is to have
\be
\frac{\del \pi_K}{\del \phi(m)} = 0
\nc
m \geq 2
\, ,
\ee
in which case $\pi_K = \pi_K(t)$.  Since $\pi_K$ only appears in the action through a term $\gradt_i \pi_K$ inside $\Ht_i$, a purely time-dependent $\pi_K$ drops out of the action.  Without loss of generality, we can set $\pi_K(t) = 0$.  Using $\pi_K = 0$ and the chain rule, the strong equation~\eqref{proofstrong} becomes
\be
0
&
=
\pi_H
- \frac{2}{d} \frac{\del \pi_H}{\del \Rt} \Rt
- \sum_{n = 2}^{\infty} n \frac{\del \pi_H}{\del \phi(n)} \phi(n)
\nl
- \frac{2 (d - 1)}{d} \frac{\del^2 \pi_H}{\del \Rt^2} \gradt_k \gradt^k \Rt
- \frac{2 (d - 1)}{d} \sum_{n=2}^\infty \frac{\del^2 \pi_H}{\del \phi(n) \del \Rt} \gradt_k \gradt^k \phi(n)
\nl
- \frac{2 (d - 1)}{d} \( \gradt_k \frac{\del^2 \pi_H}{\del \Rt^2} \) \gradt^k \Rt
- \frac{2 (d - 1)}{d} \sum_{n=2}^\infty \( \gradt_k \frac{\del^2 \pi_H}{\del \phi(n) \del \Rt} \) \gradt^k \phi(n)
\,
.
\label{strongkzero}
\ee
To eliminate dependence of the right-hand-side on the scalars $\gradt_k \gradt^k \Rt$ and $\gradt_k \gradt^k \phi(n)$ without generating new constraints, their respective scalar coefficients must vanish, {\it i.e.},
\be
\frac{\del^2 \pi_H}{\del \Rt^2} = 0
\nc
\frac{\del^2 \pi_H}{\del \phi(n) \del \Rt} = 0
\nc
n \geq 2
\, .
\ee
It follows immediately that
\be
\frac{\del \pi_H}{\del \Rt} = f(t)
\, ,
\ee
where $f(t)$ is an arbitrary function of time.  Integrating once more yields
\be
\pi_H = f(t) \Rt + g(t,\phi(n))
\, ,
\label{hgtr}
\ee
where $g(t,\phi(n))$ is an arbitrary ultralocal function.  With this result, the strong equation~\eqref{strongkzero} becomes
\be
0
&
=
\frac{d-2}{d} f(t) \Rt
+ g(t,\phi(n))
- \sum_{n = 2}^{\infty} n \frac{\del g}{\del \phi(n)} \phi(n)
\,
.
\ee
To eliminate the dependence of the right-hand-side on $\Rt$, its coefficient must vanish,
\be
\frac{d-2}{d} f(t) = 0
\, .
\ee
If $d \neq 2$, then $f(t) = 0$.  From equation~\eqref{hgtr}, it follows at once that $\pi_H$ is an ultralocal function, which is the desired result.

If $d=2$, then $f(t)$ can be arbitrary, so equation~\eqref{hgtr} appears to imply that $\pi_H$ can depend on $\Rt$.  However, $\pi_H$ contributes to the action only through a term
\be
\int \ddx \pi_H = \int \ddx g(t,\phi(n))
+ f(t) \int \ddx \Rt
\, .
\ee
Since $\th = 1$, we can equally well write this as
\be
\int \ddx \pi_H = \int \ddx g(t,\phi(n))
+ f(t) \int \ddx \sqrt{\th} \Rt
\, .
\ee
When $d = 2$, the Gauss-Bonnet theorem tells us that $\sqrt{\th} \Rt$ is a total derivative, so the latter integrand does not affect the equations of motion.  Once again, we can choose $f(t) = 0$ without loss of generality; equation~\eqref{hgtr} then implies that $\pi_H$ is an ultralocal function, which is the desired result.

\subsection{Case II: $\del \pi_K/\del \phi(n) \neq 0$ for some $n$}

If $\del \pi_K/\del \phi(n) \neq 0$ for some $n$, the vanishing of the coefficients~\eqref{strongco} requires
\be
\frac{\del^2 \pi_H}{\del \Rt^2} = 0
\nc
\frac{\del^2 \pi_H}{\del \phi(n) \del \Rt}
\nc
n \geq 2
\, .
\ee
It follows immediately that
\be
\frac{\del \pi_H}{\del \Rt} = f(t)
\, ,
\ee
where $f(t)$ is an arbitrary function of time.  Integrating once more yields
\be
\pi_H = f(t) \Rt + g(t,\phi(n))
\, ,
\label{hfrt}
\ee
where $g(t,\phi(n))$ is an arbitrary ultralocal function.

If $d = 2$, the Gauss-Bonnet theorem tells us that $\Rt = \sqrt{\th} \Rt$ is a total derivative.  Since $\pi_H$ only appears in the action through $\int \ddx \pi_H$, without loss of generality we can choose $f(t) = 0$ in equation~\eqref{hfrt}.  It follows immediately that $\pi_H$ is an ultralocal function.

If $d > 2$, applying~\eqref{hfrt} to the strong equation~\eqref{proofstrong} yields
\be
0
&
=
g(t,\phi(n))
- \sum_{n = 2}^{\infty} n \frac{\del g}{\del \phi(n)} \phi(n)
+ \frac{\del \pi_K}{\del t}
+ \frac{d-2}{d} f(t) \Rt
\nl
- \frac{1}{d} f(t) \Rt \sum_{m = 3}^{\infty} m \frac{\del \pi_K}{\del \phi(m)} \phi(m-1)
+ f(t) \sum_{m = 2}^{\infty} m \frac{\del \pi_K}{\del \phi(m)} \Pi(m-1)_{j k} \Rt^{j k}
\,
.
\label{proofhrstrong}
\ee
To eliminate dependence of the right-hand-side on the phase space scalars
\be
\Pi(m-1)_{j k} \Rt^{j k}
\nc
m \geq 2
\ee
without generating new constraints, the corresponding scalar coefficients
\be
f(t) m \frac{\del \pi_K}{\del \phi(m)}
\nc
m \geq 2
\ee
must vanish.  Since $\frac{\del \pi_K}{\del \phi(m)} \neq 0$ for some $m$, the function $f(t)$ must vanish.  Equation~\eqref{hfrt} then implies that $\pi_H$ is ultralocal.  This concludes the proof that if $\pi_K$ is ultralocal, $\pi_H$ must be also be ultralocal.

\end{document}